\documentclass[journal]{IEEEtran}

\usepackage{cite}
\usepackage{amsmath}
\usepackage{amsthm}
\usepackage{algorithm}
\usepackage{algorithmic}
\usepackage{array}
\usepackage{url}
\usepackage{color}
\usepackage{amssymb}
\usepackage{graphicx}
\usepackage{balance}

\newtheorem{theorem}{Theorem}
\newtheorem{lemma}{Lemma}
\newtheorem{definition}{Definition}
\newtheorem{corollary}{Corollary}
\newtheorem{example}{Example}

\newtheorem{remark}{Remark}

\newcommand{\off}[1]{}

\usepackage[bottom]{footmisc}

\balance

\begin{document}

\title{Successive Refinement in Large-Scale Computation: Advancing Model Inference Applications}

\author{Homa~Esfahanizadeh, Alejandro Cohen, Shlomo Shamai (Shitz), Muriel M\'edard\vspace{-0.5cm}
\thanks{H.~Esfahanizadeh is with Nokia Bell Labs, Murray Hill, NJ 07974 USA (email: homa.esfahanizadeh@nokia.com). A.~Cohen and S.~Shamai are with the Electrical and Computer Engineering Department, Technion - Israel Institute of Technology, Israel, (e-mails: \{alecohen,sshlomo\}@technion.ac.il). M.~Medard is with the Research Laboratory of Electronics (RLE), Massachusetts Institute of Technology (MIT), Cambridge, MA 02139 USA  (e-mail: medard@mit.edu). Parts of contributions of this paper were presented in IEEE International Conference on Cloud Networking (CloudNet), 2022 \cite{EsfahanizadehCloudnet2022}.}}


\maketitle

\begin{abstract}
Modern computationally-intensive applications often operate under time constraints, necessitating acceleration methods and distribution of computational workloads across multiple entities. However, the outcome is either achieved within the desired timeline or not, and in the latter case, valuable resources are wasted. In this paper, we introduce solutions for layered-resolution computation. These solutions allow lower-resolution results to be obtained at an earlier stage than the final result. This innovation notably enhances the deadline-based systems, as if a computational job is terminated due to time constraints, an approximate version of the final result can still be generated. Moreover, in certain operational regimes, a high-resolution result might be unnecessary, because the low-resolution result may already deviate significantly from the decision threshold, for example in AI-based decision-making systems. Therefore, operators can decide whether higher resolution is needed or not based on intermediate results, enabling computations with adaptive resolution. We present our framework for two critical and computationally demanding jobs: distributed matrix multiplication (linear) and model inference in machine learning (nonlinear). Our theoretical and empirical results demonstrate that the execution delay for the first resolution is significantly shorter than that for the final resolution, while maintaining overall complexity comparable to the conventional one-shot approach. Our experiments further illustrate how the layering feature increases the likelihood of meeting deadlines and enables adaptability and transparency in massive, large-scale computations.
\end{abstract}


\IEEEpeerreviewmaketitle

\section{Introduction}
\IEEEPARstart{T}{his} paper proposes a novel \textit{layered-resolution} distributed computation, where an approximated result can be obtained at an earlier stage than when releasing the final result is possible, offering multiple delay factors -- one per resolution. To highlight the importance of this added feature, consider a system with a deadline, where if the computational time of a job exceeds a threshold, it will be terminated. In such a setting, without layering, all computational resources that are spent on the terminated job are wasted. However, using our scheme, an approximation of the final result can still be released for the terminated job. Another critical application is in the decision-making process, where a high-resolution result is only needed in certain regimes of operation (e.g., near a decision threshold). In such settings, depending on the low-resolution result, the operator can decide whether to perform a resolution upgrade or not, reducing the overall system cost.\\
\vspace{-0.02cm}
Having computational responses with a small delay has a critical importance and is indeed one of the main motivations behind using parallelism and distributed solutions. The previous techniques to lower system delays in a distributed computation mainly fall into two categories: The first one is introducing redundant computations to mitigate the problem of slow workers (stragglers), known as distributed coded computation \cite{Ng2020ASO}. According to this technique, the computational load is encoded into several smaller tasks, including some redundancy, and a fusion node is able to obtain the final result of a computational job upon receiving a subset of the task results. Distributed coded computation has been studied for a variety of problems, e.g., matrix multiplication \cite{8002642,8437852,lee2017high,suh2017matrix,baharav2018straggler,yu2017polynomialn,Dutta2020,mallick2019fast,NIPS2017_e6c2dc3d,9785638}, gradient descent algorithm \cite{raviv2020gradient,dutta2019short,tandon2017gradient}, data shuffling, convolution, and fast Fourier transform \cite{song2019pliable,attia2019near,dutta2017coded,yang2016fault,jeong2018masterless,yu2017coded}, etc. The second category is utilizing scheduling to maximize the resource utilization \cite{Reisizadeh,Avestimehr}. According to this technique, nonuniform load balancing is considered to distribute the computational load among workers considering their various capabilities. Recently, joint coding and scheduling was introduced to reduce end-to-end execution delay for a system with heterogeneity and stochastic response \cite{cohen2021stream,HomaInfocom}. The idea is to close the gap between distribution response time of workers to their assigned coded tasks, via load balancing.

This paper opens a new dimension for having low-delay distributed computation, allowing approximations of the final result to be released overtime rather than having a one-shot final result. To the best of our knowledge, the previous distributed solutions, while reducing delays and uncertainties, consider only a single delay factor which indicates the time it takes from the job arrival until when the fusion node receives enough task results to release the final result. Layered resolution strategies have been rarely investigated in the realm of distributed computation which is the main scope of this paper. We note that for data communication, layering has been already considered to produce a refinement of a source
description in a successive manner \cite{SuccessRefine,SuccessRefine2,SuccessRefineComm,Successive_Refinement_Images_ML}. For example, in the broadcast communication, the layering can offer multi-level delay or multi-level quality to the users \cite{tajer2021broadcast,nikbakht2019mixed,nikbakht2020multiplexing,cohen2022broadcast,LimaMedard,KimMedard,VARIABLERATE_Shlomo2010}, and the concept also appears in the classical source coding problems, where deliveries with variable distortion are studied \cite{berger1996ceo,chen2008robust,MinimaxRateDistortion,Lossy_Source_Coding_LinZhou}. The successive refinement does not need to be limited to classical
source coding problems, where notions such as mean square errors are used to measure the distortion. In fact, the distortion can be considered as various measures for quality, e.g., image quality  in signal processing. In this paper, the distortion shows the quality of the computational result for a linear or nonlinear function, and thus we provide semantic ways for multi-resolution solutions.

There exists some classic iterative algorithms that naturally offer intermediate approximations while converging to the final high-resolution result. For example, Newton's method produces successively better approximations to the roots of a function. However, there is no natural layering solution for most massive computational tasks that frequently appear in the emerging time-sensitive applications. In this paper, we consider two models for the computational job. The first model is serving a stream of matrix-vector multiplication jobs, where the jobs arrive at the queue of a master node and are split among worker nodes. The goal is to reduce the end-to-end delays, while meeting certain deadlines. The second one is inference on deep neural network. Neural networks play a significant role in many classification tasks, and thus offering intermediate results enhances the decision-making process for many time-sensitive applications such as for autonomous vehicles or in military applications. Moreover, deep neural networks can approximate a wide variety of interesting functions, see the universal approximation theorem  \cite{HornikUniversal1989}.

According to our simulation results for the streaming distributed multiplication, the average delay of the first (lowest) resolution is notably lower than the average delay of the one-shot computation. Moreover, the success rate, i.e., the ratio of the number jobs finished before a deadline to the total number of jobs, is almost one for the first resolution, in a setting where computing the final resolution is terminated quite often. In the machine learning (ML) inference task, we demonstrate a scenario where less then $79\%$ of samples require the third resolution upgrade, and only $12\%$ and $3\%$  require the forth and fifth resolution upgrades, respectively. Our result show that the predictive utility of such an adaptive mechanism (in terms of the ROC AUC) is very close to the case where each sample goes through all the resolution upgrades.

The rest of this paper is organized as follows: In Section~\ref{sec:Problem_Setting}, we mathematically formulate the problem of computation with layered resolution and devise several necesary conditions. In Section~\ref{sec:method}, we present layering solutions for linear and nonlinear functions. In Section~\ref{sec:application_linear}, a streaming computational application is presented, benefiting from the layering solution for linear functions. In Section~\ref{sec:application_nonlinear}, an ML classification with adaptive resolution is presented that benefits from the layering solution for nonlinear functions. Section~\ref{sec:discussion} is dedicated to discussions and future work. All long proofs appear in the appendix.

\section{Problem Setting}\label{sec:Problem_Setting}
Consider a function $f:\mathbb{R}^n\rightarrow \mathbb{R}^m$, with complexity $\Theta(f)$, and arbitrary input $X\in \mathbb{R}^n$. The problem of computing $f(X)$ with $R$ layers of resolution is to enable approximations of the result to be successively released overtime with a minimal additional cost. More formally, the problem is to identify a set of bi-variate functions $\{f_1,\dots,f_R\}$, called \textit{resolution upgrades}: Each function $f_r(X,\Omega_{i-1})$ takes two inputs, $X$ and the previous approximation of $f(X)$ denoted with $\Omega_{r-1}$, and it produces a refined approximation of $f(X)$. In fact, $\Omega_{r}\triangleq f_r(X,\Omega_{r-1})$ is the $r$-th resolution upgrade of $f(X)$, $r\in\{1,\dots,R\}$, and $\Omega_0\triangleq[0]_{m}$ is an initial estimation. By definition, the order of these resolution upgrades matters. The first upgrade needs to be done first so that the second upgrade can increase the precision of computations. The computation of $f(X)$ without layering is called the \textit{one-shot computation}.

A valid solution for the described problem is called a \textit{layering strategy} in this paper, and it is required to meet the following necessary conditions:
\begin{enumerate}
    \item \textbf{[Precision Requirement]}
    Each resolution upgrade helps towards a finer approximation of the outcome and the final result must be close to the one-shot result, i.e.,
    \begin{equation}\label{eq:cond1}
        |f(X)-\Omega_{r}|< \delta(r),\;\; \forall r\in\{1,\dots,R\},
    \end{equation}
    such that $\delta(r)$ is non-increasing in $r$, and $\delta(R)<\delta$ for a desired $\delta$.
    \item \textbf{[Complexity Requirement]}
    Each resolution upgrade must have a complexity lower than the one-shot computational complexity, i.e.,
    \begin{equation}\label{eq:cond2}
    \Theta(f_r) < \Theta(f)- \epsilon(r),\;\; \forall r\in\{1,\dots,R\},
    \end{equation}
    for a desired $\epsilon(r)\in\mathbb{R}^+$.
\end{enumerate}
A superior layering strategy offers smaller value for $\delta(r)$ and larger value for $\epsilon(r)$, since it enables the intermediate results faster with less overall precision degradation.

We next describe the mathematical notations used in the rest of the paper: Matrices and vectors are represented with upper-case letters, and scalars are represented with lower-case letters. The function $|.|$ and $\lfloor . \rfloor$ and $sgn(.)$ denote the point-wise absolute, floor, and sign, respectively. The subscript for a matrix or vector indicates its dimension. The all-one matrix and all-zero matrix are shown as  $[1]_{m\times n}$ and $[0]_{m\times n}$. The notation $A(i)$ and $A(i,j)$ are used to denote a specific element of a vector or matrix. For addressing a specific row or column of a matrix, we use $A(i,:)$ and $A(:,j)$, respectively.

\section{Designing Layering Strategies}\label{sec:method}

In this section, we derive a mathematical framework for computation of a function $f(X)$ with several layers of resolution. We first present our layering strategy for the case that the computational job is a linear transformation, in Section~\ref{sec:linear}. Then, we present our method for nonlinear functions, in form of evaluation of a deep neural network with finite number of hidden layers, in Section~\ref{sec:nonlinear}.

\subsection{Layered Evaluation of Linear Functions}\label{sec:linear}

Consider $f(X)=WX$ to be the computational job that we intend to present a layering strategy for, where $X\in\mathbb{R}^n$ and $W\in\mathbb{R}^{m\times n}$. Both $X$ and $W$ have finite elements\footnote{There exist $w_{\text{max}}$ and $x_{\text{max}}$, such that $|x_j| \leq x_{\text{max}}$ and $|w_{i,j}| \leq w_{\text{max}}$ for every $i\in\{1,\dots,m\}$ and $j\in\{1,\dots,n\}$.}. We first present necessary definitions and preliminary results that will be used in our layering strategy.


\begin{definition}[Partitioning Vector]\label{def:quantization}
    The partitioning vector $P=[P_0,\dots,P_d]$ is a set of distinct integers in the descending order, which is used to approximate a fractional number $\alpha$ such that $|\alpha|< 2^{P_0}$, as follows
    \begin{equation*}
        \Tilde{\alpha}= s_\alpha\sum_{i=1}^{d}\alpha_i 2^{P_i}.
    \end{equation*}
    Here, $\alpha_i\in\{0,\dots,2^{P_{i-1}-P_{i}}-1\}$ $\forall i\in\{1,\dots,d\}$, $s_\alpha\in\{+1,-1\}$ represents the sign of $\alpha$, and $|\alpha-\Tilde{\alpha}|<2^{P_d}$.
\end{definition}

Definition~\ref{def:quantization} states that the contribution of each number in $\{\alpha_1,\dots,\alpha_d\}$ to $\Tilde{\alpha}$ is the value of the number multiplied by a factor determined by the position of the number. This representation has a direct connection to the positional system and the binary representation of a real number. Each non-negative real number $\alpha$ can be described as $\sum_{k=-\infty}^{\infty}b_k 2^k$, where $b_k$ is a bit. Assuming $\alpha< 2^{P_0}$, we have $\alpha=\sum_{k=-\infty}^{P_0-1}b_k 2^k$. By ignoring the bits that have positional values smaller than $P_d$, we obtain the approximation $\Tilde{\alpha}=\sum_{k=P_d}^{P_0-1}b_k 2^k$. Then,
\begin{equation*}
    \alpha_{i}=\sum_{k=0}^{P_{i-1}-P_{i}-1}b_{k+P_i}2^{k},
\end{equation*}
for $i\in\{1,\dots,d\}$. The approximation is precise (i.e., $|\alpha-\Tilde{\alpha}|=0$), if $b_k=0\;\;\forall k<P_d$.

\begin{example}
    For $\alpha=-1.625$ and $P_\alpha=[1,-1,-3]$, we have $\alpha_1=3$, $\alpha_2=1$, and $s_\alpha=-1$. In this case, $|\alpha-\Tilde{\alpha}|=0$.
\end{example}


We then start with a simple example, showing how multiplication of two scalars can be performed in multiple layers of resolution. Then, we extend the idea to linear transformations in a multi-dimensional space. Let $\alpha$, $\beta$ be two scalars, described in the following form according to the partitioning vectors $P_\alpha=[P_0,\dots,P_d]$ and $P_\beta=[Q_0,\dots,Q_d]$, respectively,
\begin{equation*}
    \alpha\simeq s_\alpha\sum_{i=1}^{d}\alpha_i 2^{P_i},\;\;\beta\simeq s_\beta\sum_{i=1}^{d}\beta_i 2^{Q_i}.
\end{equation*}
Then,
\begin{equation*}
    \alpha\beta\simeq s_\alpha s_\beta\sum_{i=1}^{d}\sum_{j=1}^{d}\alpha_i\beta_j 2^{P_i+Q_j}.
\end{equation*}

The function evaluation can be done in $R=d^2$ levels of resolution: At each level, a multiplication $\alpha_i\beta_j$ is performed, where $(i,j)\in\{1,\dots,d\}\times\{1,\dots,d\}$, with priories given to those with higher $P_i+Q_j$ (corresponding to the most significant parts of the computational result). Then, the result is added with appropriate positional shift, to refine the approximation. This process is more-formally described below:

\begin{definition}\label{def:Gamma_vec}
    We define $\Gamma(P_\alpha,P_\beta)=[(\gamma_i(1),\gamma_i(2))]_{i=1}^{d^2}$ to be an ordered version of tuples in $\{1,\dots,d\}^2$, such that if $i\prec j$, $P_{\gamma_i(1)}+Q_{\gamma_i(2)}\geq P_{\gamma_j(1)}+Q_{\gamma_j(2)}$.
\end{definition}

Given $P_\alpha$ and $P_\beta$, we obtain $\Gamma(P_\alpha,P_\beta)$ according to Definition~\ref{def:Gamma_vec}. Then, the $r$-th resolution of computing $\alpha\beta$, where $r\in\{1,\dots,d^2\}$, is:
\begin{equation*}
\begin{split}
    \Omega_r&=s_\alpha s_\beta\sum_{i=1}^{r}\alpha_{\gamma_i(1)}\beta_{\gamma_i(2)} 2^{P_{\gamma_i(1)}+Q_{\gamma_i(2)}}\\
    &=\Omega_{r-1}+\left(s_\alpha s_\beta \alpha_{\gamma_r(1)}\beta_{\gamma_r(2)} 2^{P_{\gamma_r(1)}+Q_{\gamma_r(2)}}\right),
\end{split}
\end{equation*}
and $\Omega_0=0$.

\begin{example}
    Consider $\alpha=-1.625$ and $\beta=13.125$. For two partitioning vectors $P_\alpha=[1,-1,-3]$ and $P_\beta=[4,0,-3]$, we have $\alpha_1=3$, $\alpha_2=1$, $\beta_1=13$, $\beta_2=1$. Besides, $s_\alpha=-1$ and $s_\beta=1$. According to Definition~\ref{def:Gamma_vec},
    $$\Gamma=[(1,1),(2,1),(1,2),(2,2)].$$
    Therefore,
    \begin{equation*}
    \small
        \begin{split}
            \Omega_1&{=}(3\times 13)2^{-1+0}=19.5\\
            \Omega_2&{=}\Omega_1+(1\times 13)2^{-3+0}=19.5+1.625=21.125\\
            \Omega_3&{=}\Omega_2+(3\times 1)2^{-1-3}=21.125+0.1875=21.3125\\
            \Omega_4&{=}\Omega_3+(1\times 1)2^{-3-3}=21.3125+0.015625=21.328125\\
        \end{split}
    \end{equation*}
\end{example}

Finally, we extend the layering mechanism for linear transformations in a high-dimensional space, i.e., $f(X)=WX$. The partitioning of a matrix $A$ using the partitioning vector $P_A=[P_0,\dots,P_d]$ is,
\begin{equation}\label{eq:partition_matrix}
    A\simeq \sum_{i=1}^{d}A_i 2^{P_i}.
\end{equation}
The above representations are obtained by applying Definition~\ref{def:quantization} to each element of $A$, with a difference that the sign of each element of $A$ also appears in each element of $A_i$. Therefore, each element of $A_i$ belongs to $\{-2^{P_{i-1}-P_{i}}+1,\dots 0,\dots,2^{P_{i-1}-P_{i}}-1\}$, where $i\in\{1,\dots,d\}$. Algorithm~\ref{alg:partition} describes a pseudo code for partitioning a matrix/vector $A$ given its partitioning vector $P_A$. At first, the sign of each element of $A$ is stored in $S_A$ (line~3). Then, the absolute value of each element is taken and is divided by $2^{P_0}$ (line~4). By this time, the elements in $A$ must be less than $1$ if $A<2^{P_0}$, and if not line~5 removes the integer parts. Then, starting from $i=1$, each element is multiplied by $2^{P_{i-1}-P_{i}}$. The integer part is multiplied element-wise by $S_A$ to obtain $A_i$, and the fractional number is extracted for next iteration (lines 6-10).

\begin{algorithm}[t]\small
\caption{$partition(A,P_A)$}
\label{alg:partition}
\begin{algorithmic}[1]
\STATE \textbf{Input:} $A$ and $P_A=[P_0,\dots,P_d]$
\STATE \textbf{Output:} $[A_1,\dots,A_d]$
\STATE $S_A = sgn(A)$
\STATE $A = |A| / 2^{P_0}$
\STATE $A = A  - \lfloor A  \rfloor$
\FOR{$i\in[1,\dots,d]$}
\STATE $A=A 2^{P_{i-1}-P_{i}}$
\STATE $A_i = S_A\odot \lfloor P_A \rfloor  $
\STATE $A = A  - \lfloor A  \rfloor$
\ENDFOR
\end{algorithmic}
\end{algorithm}

We consider the following representations for $W$ and $X$, given the partitioning vectors $P_W=[P_0,\dots,P_d]$ and $P_X=[Q_0,\dots,Q_d]$, using Algorithm~\ref{alg:partition}.
\begin{equation*}
    W\simeq \sum_{i=1}^{d}W_i 2^{P_i},\;\;X\simeq \sum_{i=1}^{d}X_i 2^{Q_i}.
\end{equation*}
Thus,
\begin{equation*}
    f(X)=WX\simeq\sum_{i=1}^{d}\sum_{j=1}^{d}W_iX_j 2^{P_i+Q_j}.
\end{equation*}

\begin{definition}\label{def:layering_linear}
    The $r$-th resolution upgrade of $f(X)=WX$ is obtained as follows, where $r\in\{1,\dots,R\}$ and $R=d^2$.
    \begin{equation*}
    \begin{split}
    \Omega_r&=f_r(X,\Omega_{r-1})=\sum_{i=1}^{r}W_{\gamma_i(1)}X_{\gamma_i(2)} 2^{P_{\gamma_i(1)}+Q_{\gamma_i(2)}}\\
    &=\Omega_{r-1}+W_{\gamma_r(1)}X_{\gamma_r(2)} 2^{P_{\gamma_r(1)}+Q_{\gamma_r(2)}}.
    \end{split}
\end{equation*}
The values for $[(\gamma_i(1),\gamma_i(2))]_{i=1}^{R}$ are identified according to Definition~\ref{def:Gamma_vec}.
\end{definition}

\begin{theorem}\label{theorem:valid_linear1}
For the layering strategy in Definition~\ref{def:layering_linear}, we have
    \begin{equation*}
        |f(X)-\Omega_{r}|< \delta(r)\;\; \forall r\in\{1,\dots,R\},
    \end{equation*}
where $\delta(r)=\delta+(R-r)2^{P_{\gamma_r(1)}+Q_{\gamma_r(2)}}$ is non-increasing in $r\in\{1,\dots,R\}$ and $\delta=2^{P_0+Q_d}+2^{P_d+Q_0}+2^{P_d+Q_d}$.
\end{theorem}

\begin{remark}\label{remark:on_appending}
    The layering strategy in Definition~\ref{def:layering_linear}, can be readily applied to linear transformations in form of $f(X)=WX+B$. For this, we define $f(X)=W'X'$, where
    \begin{equation*}
        W'=[W\;B]\;\;\text{and}\;\;X'=\left[\begin{array}{c}
             X  \\
             1
        \end{array}\right].
    \end{equation*}
\end{remark}

\begin{remark}\label{remark:on_linear}
Although the presented layering strategy is defined based on the binary field, the solution can be easily extended to other bases as well.
\end{remark}

Theorem~\ref{theorem:valid_linear1} states that the presented solution  for linear functions meets the first requirement of a valid layering strategy (precision requirement). The second condition requires the complexity of each resolution upgrade to be smaller than the one for one-shot computation. In this paper, we consider the computational complexity of multiplication of two integers with $m$ bits and $n$ bits is $\Theta(f)=mn$.\footnote{It has recently been shown that multiplication has a theoretical time complexity of $O(m \log n)$, assuming $m>n$ \cite{Harvey_multiplication_complexity_2021}. However, algorithms in practice have a time complexity of $O(mn)$. Nevertheless, the analysis can be adjusted based on the computational complexity of multiplication.} We further consider that multiplication of a $u\times v$ matrix with $m$-bit elements and a $v\times 1$ vector with $n$-bits elements has computational complexity $\Theta(f)=uvmn$, neglecting the cost of addition.

\begin{theorem}\label{theorem:valid_linear2}
For the layering strategy in Definition~\ref{def:layering_linear}, we have
    \begin{equation*}
    \sum_{i=1}^R\Theta(f_i)\approx \Theta(f).
    \end{equation*}
\end{theorem}
Theorem~\ref{theorem:valid_linear2} implies that $\Theta(f_r)=\Theta(f)-\epsilon(r)$, where $\epsilon(r)=\sum_{i=1,i\neq r}^R\Theta(f_i)>0$ for $R>1$, and thus the second condition of a valid layering strategy (complexity requirement) is met.

\subsection{Layered Evaluation of nonlinear Functions}\label{sec:nonlinear}

\begin{figure}
    \centering
    \includegraphics[width=0.99\columnwidth]{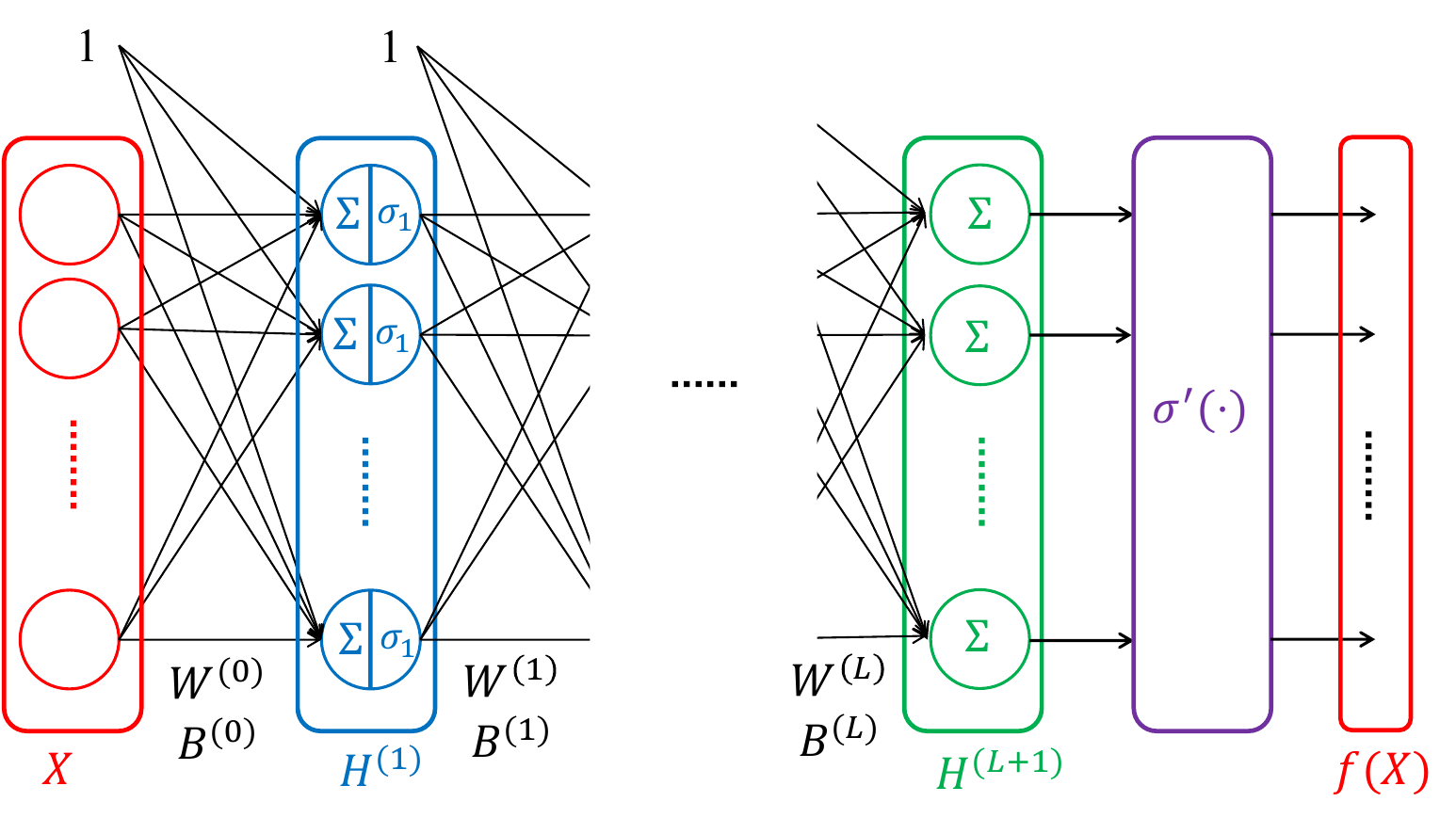}\vspace{-0.3cm}
    \caption{Architecture of a deep neural network with $L$ hidden layers. The $i$-th hidden layer $H^{(i)}$ is identified via a linear transformation $\{W_{i-1},B_{i-1}\}$, followed by a point-wise nonlinear function $\sigma_{i}$; except for the last layer that is followed by a nonlinear function $\sigma'$ that is not necessarily point-wise. Each edge is associated with a weight that is described by the linear transformation.}
    \label{fig:deep_nn}
\end{figure}

\begin{figure}
    \centering
    \begin{tabular}{cc}
    \includegraphics[height=0.29\columnwidth]{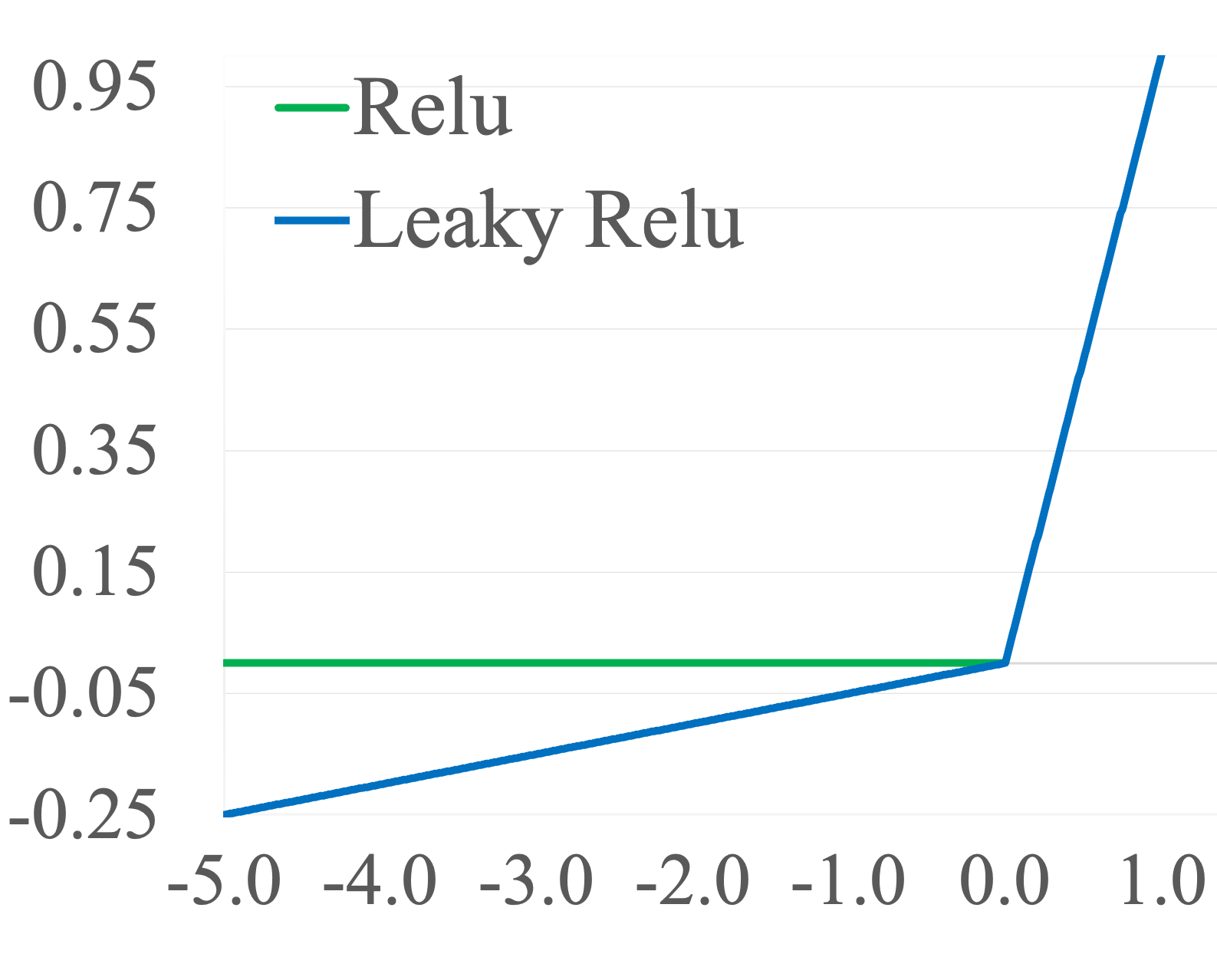}&
    \includegraphics[height=0.29\columnwidth]{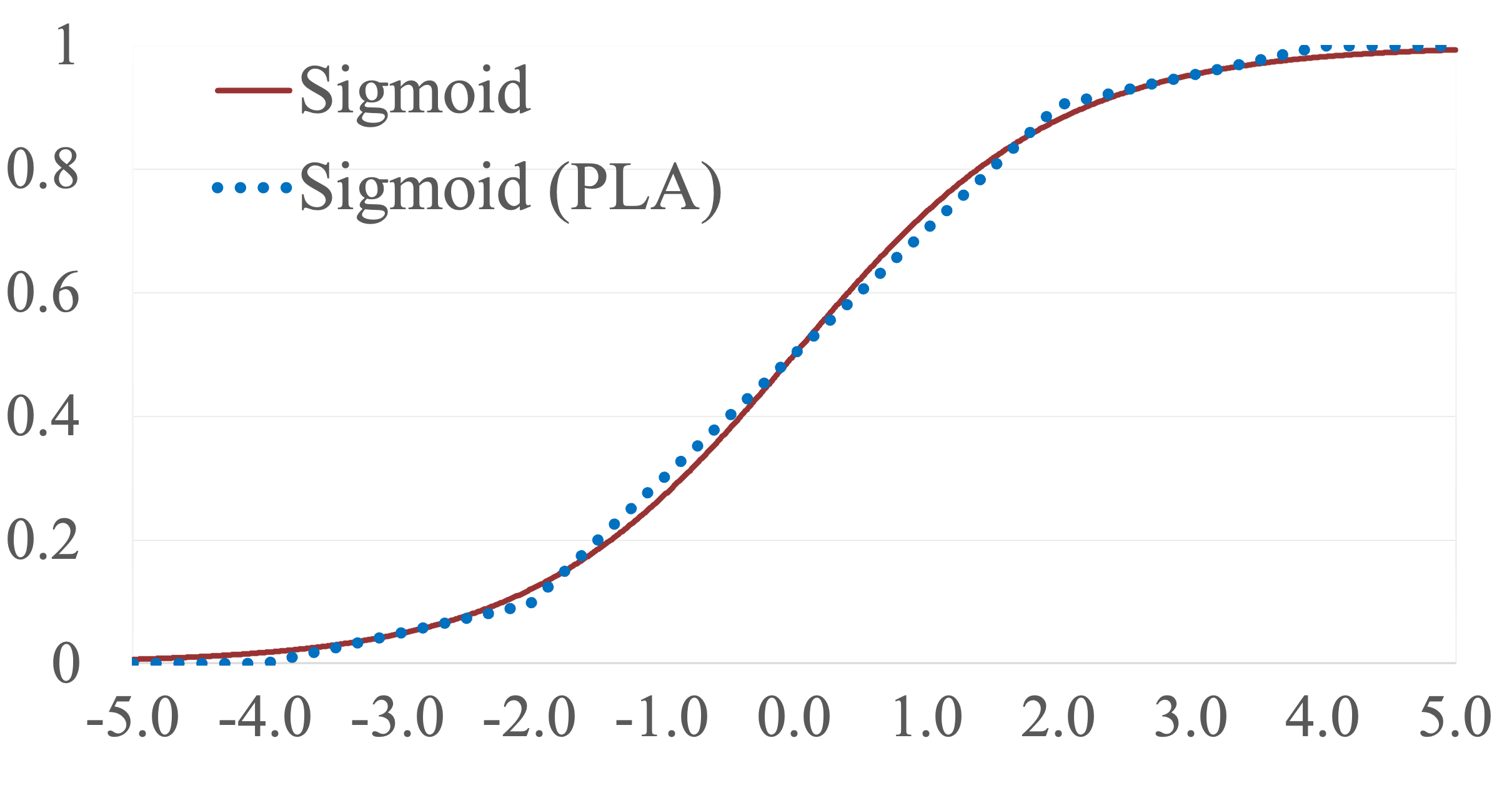}\\
    (a)&(b)
    \end{tabular}
    \caption{Examples of point-wise activation functions: (a) piece-wise linear functions Relu $\sigma(a)=\max(a,0)$ and Leaky Relu $\sigma(a)=\max(a,\beta a)$, $\beta=0.05$ here; (b) Sigmoid  $\sigma(a)=1/(1+\exp(-a))$ and its piece-wise linear approximation (PLA).}
    \label{fig:activations}
    \vspace{-0.3cm}
\end{figure}

A deep neural network is defined as composition of several linear and nonlinear transformations, which is the computational job that we intend to present a layering strategy for,
$$f(X)=\sigma'(W^{(L)}\sigma_{L}(\dots \sigma_1(W^{(0)}X+B^{(0)})\dots)+B^{(L)}).$$
Here, $X\in\mathbb{R}^{n_0}$, $W^{(l)}\in \mathbb{R}^{n_{l+1}\times n_{l}}$, and $B^{(l)}\in \mathbb{R}^{n_{l+1}}$, all having finite elements and $l\in\{0,\dots,L\}$, see Fig.~\ref{fig:deep_nn}. Besides, $\sigma_i:\mathbb{R}\rightarrow \mathbb{R}$, where $i\in\{1,\dots,L\}$, are point-wise nonlinear functions, known as activation functions. Several common activation functions are depicted in Fig.~\ref{fig:activations}. Function $\sigma':\mathbb{R}^m\rightarrow \mathbb{R}^m$ acts on the results of the last linear transformation and outputs the final result, e.g., softmax converts a vector of $m$ real numbers into a probability distribution of $m$ possible outcomes. By construction, we have $n_0=n$ and $n_{L+1}=m$. Using Remark~\ref{remark:on_linear}, we can also write a neural network computation as follows,
\begin{equation}\label{eq:deep_nn_mod}
    f(X)=\sigma'(W^{(L)}\sigma_{L}(\dots W^{(1)}\sigma_1(W^{(0)}X)\dots)).
\end{equation}

In this paper, we assume the nonlinear activations are either piece-wise linear, or can be approximated by piece-wise linear functions, see Fig.~\ref{fig:activations}. This simplification enables us to build an efficient layering strategy for evaluation of a neural network, while marginally effecting the evaluation precision and the overall complexity. We define the following intermediate variable per hidden layer of a neural network. For $l\in\{1,\dots,L+1\}$,
\begin{equation}\label{eq:hiddeneq}
    H^{(l)}= W^{(l-1)}\sigma_{l-1}(H^{(l-1)}),
\end{equation}
where $H^{(0)}=X$ and $\sigma_0(a)=a$. Thus, $f(X)=\sigma'(H^{(L+1)})$, see Fig.~\ref{fig:deep_nn}. Similar to the layered-resolution computation of linear transformations, we consider for $X$ and $W^{(l)}$, partitioning vectors $P_X$ and $P_{W,l}$, consists of integers in descending order.\footnote{These vectors are the system designer's choice, and an appropriate choice depends on  the distributions of their  elements. We will present some heuristic approaches in our results in Section~\ref{sec:application_nonlinear}.}

\begin{lemma}\label{lemma_nonlinearity}
    For a piece-wise linear function $\sigma(x)=\rho_i x +\nu_i$, $x\in \Pi _i$, we have
\begin{equation*}
    \sigma(t + \Delta t)=\sigma(t)+\sigma^t(\Delta t),
\end{equation*}
where
\begin{equation*}
    \sigma^t(\Delta t){=}
    \begin{cases}
        \rho_i \Delta t& t,t+\Delta t\in \Pi_i\\
        (\rho_i-\rho_j)t + \rho_i \Delta t+\nu_i - \nu_j&t\in\Pi_j,t+\Delta \in \Pi_i\\
    \end{cases}
\end{equation*}
\end{lemma}
\begin{proof}
When both $t,t+\Delta t\in \Pi_i$,
\begin{equation*}
\sigma(t+\Delta t)=\rho_i(t+\Delta t)+\nu_i=\sigma(t)+\rho_i\Delta t.
\end{equation*}
When $t\in\Pi_j,t+\Delta \in \Pi_i$,
\begin{equation*}
\begin{split}
\sigma(t+\Delta t)&=\rho_i(t+\Delta t)+\nu_i\\
&=\sigma(t)-\rho_j t - \nu_j+\rho_i(t+\Delta t)+\nu_i\\
&=\sigma(t)+(\rho_i-\rho_j)t + \rho_i \Delta t+\nu_i - \nu_j
\end{split}
\end{equation*}
\end{proof}

\begin{corollary}
    For the Relu activation function $\sigma(t)=\max(x,0)$, we have
\begin{equation*}
\sigma_t(\Delta t)=\begin{cases}
        \Delta t& t\geq 0,t+\Delta t \geq 0 \\
        -t & t\geq 0,t+\Delta t < 0 \\
        t+\Delta t & t< 0,t+\Delta t \geq 0 \\
        0 & t< 0,t+\Delta t < 0
    \end{cases}.
\end{equation*}
\end{corollary}

Now, let assume the evaluation of a neural network was performed using approximations of $X$ and $W^{(l)}$, $l\in\{0,\dots,L\}$. We want to refine the evaluation using the updated values $X+\Delta X$ and $W^{(l)}+\Delta W^{(l)}$.

\begin{lemma}\label{lemma:distortion}
    Given an evaluation of a neural network as,
    $$\{H^{(l)}:l\in\{1,\dots,L+1\}\},$$
    and a set of updates for the network parameters as
    $$\{\Delta X, \Delta W^{(0)},\dots,\Delta W^{(L)}\},$$
    the updates can be obtained as follows:
\begin{flalign*}
    &\Delta H^{(0)}=\Delta X\\
    &\Delta H^{(l)}=W^{(l-1)}\sigma_{l-1}^{H^{(l-1)}}(\Delta H^{(l-1)})\\
    &+\Delta W^{(l-1)}(\sigma_{l-1}(H^{(l-1)})+\Delta W^{(l-1)}\sigma_{l-1}^{H^{(l-1)}}(\Delta H^{(l-1)}))\\
    &\Omega= \sigma'(H^{(L+1)}+\Delta H^{(L+1)}).
\end{flalign*}
\end{lemma}
\begin{proof}
Let denote the updated values of the hidden layers as $H^{(l)}+\Delta H^{(l)}$, for
$l\in \{0,\dots,L+1\}$. Because $H^{(0)}=X$, we have $\Delta H^{(0)}=\Delta X$. According to (\ref{eq:hiddeneq}),
\begin{equation*}
        H^{(l)}+\Delta H^{(l)}=(W^{(l-1)}+\Delta W^{(l-1)})\sigma_{l-1}(H^{(l-1)}+\Delta H^{(l-1)}),
\end{equation*}
for $l\in\{1,\dots,L+1\}$. Using Lemma~\ref{lemma_nonlinearity}, we have
\begin{equation*}
\begin{split}
        &H^{(l)}+\Delta H^{(l)}=\\
        &(W^{(l-1)}+\Delta W^{(l-1)})(\sigma_{l-1}(H^{(l-1)})+\sigma_{l-1}^{H^{(l-1)}}(\Delta H^{(l-1)})),
\end{split}
\end{equation*}
Since $H^{(l)}=W^{(l-1)}\sigma_{l-1}(H^{(l-1)})$, we have
\begin{equation*}
\begin{split}
\Delta& H^{(l)}=W^{(l-1)}\sigma_{l-1}^{H^{(l-1)}}(\Delta H^{(l-1)})\\
    &+\Delta W^{(l-1)}\sigma_{l-1}(H^{(l-1)})+\Delta W^{(l-1)}\sigma_{l-1}^{H^{(l-1)}}(\Delta H^{(l-1)}).
\end{split}
\end{equation*}
Finally, $\sigma'(H^{(L+1)}+\Delta H^{(L+1)})$.
\end{proof}

\begin{algorithm}[t]\small
\caption{Layered Evaluation of a Neural Network.}
\label{alg:main_algo}
\begin{algorithmic}[1]
\STATE \textbf{Input:} $R$, $X$, $W^{(0)}\dots,W^{(L)}$, $P_X$, $P_{W,0},\dots,P_{W,L}$, $h_\text{min}$\label{algo2:1}
\STATE \textbf{Initialize:} ${A_{0}}=\dots=A_{L}=[0]$\label{algo2:2}
\STATE \textbf{Initialize:} $H^{(0)}=\dots=H^{(L+1)}=[0]$\label{algo2:3}
\STATE $[X_1,\dots,X_R]={partition} (X,P_X)$, Algorithm~\ref{alg:partition}\label{algo2:4}
\FOR {$l\in[0,\dots,L]$}\label{algo2:5}
\STATE $[W_{l,1},\dots,W_{l,R}]={partition} (W^{(l)},P_{W,l})$, Algorithm~\ref{alg:partition}
\ENDFOR \label{algo2:6}
\FOR{$r\in [1,\dots,R]$}\label{algo2:7}
\FOR {$l\in[0,\dots,L]$}\label{algo2:8}
\STATE $\Delta A_l=W_{l,r} 2^{P_{W,l}(r)}$
\ENDFOR\label{algo2:9}
\STATE $\Delta H=X_r 2^{P_X(r)}$.\label{algo2:10}
\FOR{$l\in[0,\dots,L]$}\label{algo2:13}
\STATE $\Delta H_\sigma = \sigma_{l}^{H^{(l)}}(\Delta H)$\label{algo2:17}
\STATE $H=[(H^{(l)})^T,1]^T$, \; $\Delta H=[\Delta H_\sigma^T,0]^T$\label{algo2:12}
\STATE $H^{(l)}=H^{(l)}+\Delta H$\label{algo2:16}
\STATE $\Delta H =A_l \Delta H_\sigma+\Delta A_l \sigma_l(H) + \Delta A_l \Delta H_\sigma$\label{algo2:14}
\STATE $\Delta H = \lfloor \Delta H / 2^{h_\text{min}} \rfloor 2^{h_\text{min}}$\label{algo2:15}
\STATE $A_l = A_l + \Delta A_l$\label{algo2:20}
\ENDFOR
\STATE $H^{(L+1)}=H^{(L+1)}+\Delta H$\label{algo2:22}
\STATE $\Omega_r=\sigma'(H^{(L+1)})$\label{algo2:21}
\ENDFOR
\end{algorithmic}
\end{algorithm}
\vspace{-0.2cm}

We present our layering strategy for evaluation of a deep neural network in Algorithm~\ref{alg:main_algo}. The inputs are the full-resolution input sample and model weights, the partitioning vectors which also indicate the number of resolutions $R$, and an integer $h_\text{min}$ showing the number of fractional bits needed for the intermediate variables (line~\ref{algo2:1}). The initial weights that are considered for the model evaluation, and correspondingly, the intermediate variables are set to zero (lines~\ref{algo2:2}-\ref{algo2:3}). Using the partitioning vectors and Algorithm~\ref{alg:partition}, the input sample and the model weights are partitioned into $R$ components (lines~\ref{algo2:4}-\ref{algo2:6}). Then, for each resolution upgrade, starting from the first one, we need to identify the model updates. For this, the weight updates are first identified (lines~\ref{algo2:8}-\ref{algo2:9}). For the 0-th layer, the model update is simply the input update (line~\ref{algo2:10}). For the update of the $(l+1)$-th layer of the model, where $l\in\{0,\dots,L\}$, first, the value of $\Delta H$ obtained from the previous layer is passed through the nonlinear function $\sigma_{l}^{H^{(l)}}(.)$ (line~\ref{algo2:17}). Then, appropriate zero- and one-elements are appended, based on Remark~\ref{remark:on_appending} (line~\ref{algo2:12}). Next, the updated value of the $l$-th layer is recorded for the next resolution upgrade (line~\ref{algo2:16}), and finally, Lemma~\ref{lemma:distortion} is used to obtain the model update (line~\ref{algo2:14}). The bits of $\Delta H$ that have positional values less then $h_\text{min}$ are pruned to manage the overall computational complexity (line~\ref{algo2:15}), and the updated value is recorded for the next resolution upgrade (line~\ref{algo2:16}). After the final layer update, the refined result can be released for the resolution upgrade  (lines~\ref{algo2:22}-\ref{algo2:21}).

\begin{theorem}\label{theorem:valid_nonlinear1}
For the layering strategy in Algorithm~\ref{alg:main_algo}, we have
    \begin{equation*}
        |f(X)-\Omega_{r}|< \delta(r)\;\; \forall r\in\{1,\dots,R\},
    \end{equation*}
where,
\begin{equation*}
\begin{split}
    \delta(r)&=2^{Q_{r}}\left(J_\text{max}m\prod_{i=0}^{l-1} 2^{P_0^{(i)}}n_{i}\right) \\
    &+ \sum_{i=0}^{l-1}2^{P_r^{(i)}}\left(
    J_\text{max}m2^{h_\text{max}}n_{i}
    \prod_{j=i+1}^{l-1}2^{P_0^{(j)}}n_{j}\right),
\end{split}
\end{equation*}
and $\delta(r)$ is non-increasing in $r\in\{1,\dots,R\}$. Here, $J_\text{max}$ is the maximum value that Jacobian matrix of the final activation function can take, and $h_\text{max}$ is the maximum positional value that the bits of $|H^{(l)}|$ can take.
\end{theorem}

\begin{theorem}\label{theorem:valid_nonlinear2}
For the layering strategy in Definition~\ref{def:layering_linear}, we have $\Theta(f_r)<\Theta(f)-\epsilon(r)$, where
\begin{equation*}
\begin{split}
    &\epsilon(r)=\sum_{l=0}^{L} n_{l}n_{l-1}(h_\text{max}{-}h_\text{min})\\
    &\hspace{3cm}(0.5P_0^{(l)}{-}1.5P_{r-1}^{(l)}{+}2P_r^{(l)}{-}P_R^{(l)}),
\end{split}
\end{equation*}
ant it is positive in practice. Here, $h_\text{max}$ and $h_\text{min}$ are the maximum and minimum positional values that the bits of $|H^{(l)}|$ can take, respectively.
\end{theorem}

\section{Stream Distributed Matrix Multiplication}\label{sec:application_linear}

A distributed computational system consists of multiple workers that can run as a single system to collaboratively respond to heavy computational jobs. Distributed computing offers major benefits such as scalability, parallelism, robustness, and cost efficiency. Delay has a major importance in distributed systems, as emerging computational applications demand a timely response to their computational jobs to make real-time automatic decision, e.g., autonomous driving and smart cities. In fact, often a computational response matters only if it is received within a reasonable time.

\subsection{System Design}

\begin{figure}
    \centering
    \includegraphics[width=0.99\columnwidth]{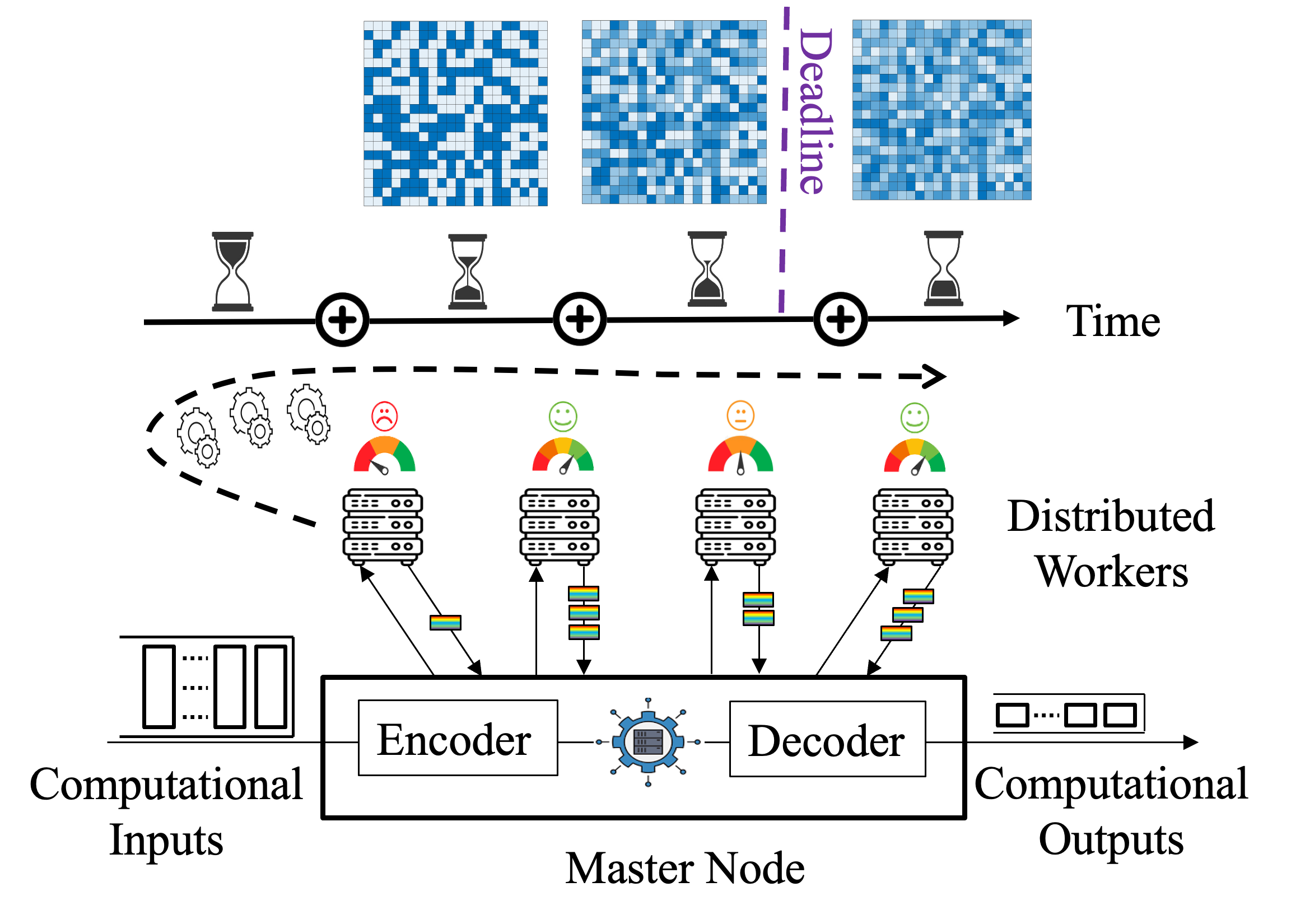}\vspace{-0.3cm}
    \caption{Stream distributed computation with successive refinement: A subset of task results are sufficient to obtain the first resolution result, and successively, the resolution is improved upon collecting more task results.}
    \label{fig:demo}
    \vspace{-0.3cm}
\end{figure}

The distributed setting we consider in this section consists of a master node and a heterogeneous cluster of $P$ workers, see Fig.~\ref{fig:demo}. The master node serves the computational jobs according to their order of arrival. It splits the computational load into several smaller tasks (including some redundant tasks), and distributes the tasks among the workers, depending on their capabilities. The workers send back the result of each computational task to the master node when ready. The fusion node can upgrade the resolution of the final result upon receiving a sufficient subset of task results. The master node removes from the system the remaining tasks from the resolved job, called purging. It then encodes and distributes the computations related to the next job in its queue (if any).

Our distributed system can be modeled with a G/G/1 queue, i.e., a system where job inter-arrival time has a general distribution and job service time (collaboratively performed by the workers) has another general distribution \cite{Bhat2015}. For simplicity, we assume the communication delays are negligible, and we target matrix-vector multiplication as our example of computational jobs. Each job is associated with a matrix
$W$ and a vector $X$, and the system need to find their multiplication $f(X)=WX$ using a distributed and multi-resolution strategy.

Our solution is based on the layering strategy described in Section~\ref{sec:linear}. The master node serves the jobs according to their order of arrival as follows:  By the means of the partitioning matrices $P_W=[P_0,\dots,P_d]$ and $P_X=[Q_0,\dots,Q_d]$, the master node obtains several matrix-vector multiplications per job, called \textit{mini-jobs}, each serving to one resolution upgrade. It is then these mini-jobs that are split into smaller task and are distributed among the workers, starting from the mini-job of the first layer. After the first resolution upgrade, the master node distributes the computations related to the next resolution upgrade of the same job or the first resolution of the next job.

We define execution delay of the $r$-th resolution of a job, denoted with $D(r)$, as the time the job arrives at the queue of the master node, until the time it is able to release the $r$-th resolution of the result. Since $D(r)$ has stochastic behaviour, we are interested in its distribution and expected value. We also occasionally consider a deadline, by which if the computational time of the job exceeds and there are other jobs in the queue, processing of the job will be terminated, and the last obtained resolution of its final result will be released by the fusion node, Fig.~\ref{fig:demo}.

Let denote with $T_{p}$ the time it takes for the $p$-th worker to compute one complete job (full resolution matrix-vector multiplication), and denote with $T_s$ the time it takes for all workers to collaboratively compute one job. The statistical parameters $E[T_p]$ and $E[T_p^2]$ are available, either by the designer or by tracking the workers' behaviour. Then, the service rate of the $p$-th worker is $1/E[T_p]$. A lower bound on $E[T_s]$ is obtained by approximating the whole system with just one worker whose service rate is summation of the service rates of all workers \cite{HomaInfocom}, i.e., $E[T_s]\geq1/{\sum_{p=1}^{P}({1}/{E[T_p]}}).$ For G/G/1 queuing model, with job inter-arrival time $T_a$, the average execution time (from job arrival to delivery and thus including the queue delay) is approximated by \cite{marchal1976approximate}:
\begin{equation}\label{eq:Kingmansformula}
    E[D]\approx E[T_s]+E[T_s]\left(\frac{\rho}{1-\rho}\frac{c_a^2+c_s^2}{2}
    \right).
\end{equation}
Here $\rho=E[T_s]/E[T_a]$, $c_a^2=(E[T_a^2]-E[T_a]^2)/E[T_a]^2$, and $c_s^2=(E[T_s^2]-E[T_s]^2)/E[T_s]^2$.
Incorporating the lower bound of $E[T_s]$ into (\ref{eq:Kingmansformula}) results in a lower bound for the average job delay for the distributed system with no layering. Here, the first part of the summation represents the average computational delay and the second part represents the average queuing delay.

When we also incorporate layering, the queuing delay is still the same, for a system with no job termination. However, the computational delay of lower layers are smaller. Let $T_s^l$ represent the time it takes for all workers to collaboratively compute the $r$-th resolution of the job. All mini-job have the same complexity. The ratio of computational complexity of the $r$-th resolution upgrade to the one-shot computation is (See proof of Theorem~\ref{theorem:valid_linear2}),
\begin{equation*}
    \frac{\Theta(f_r)}{\Theta(f)}=\frac{(P_{\gamma_r(1)-1}-P_{\gamma_r(1)})(Q_{\gamma_r(2)-1}-Q_{\gamma_r(2)})}{(P_0-P_d)(Q_0-Q_d)}.
\end{equation*}
Thus,
\begin{equation}\label{eq:LB}
\begin{split}
    &E[T_s^l]\geq \frac{\sum_{i=1}^r(P_{\gamma_i(1)-1}-P_{\gamma_i(1)})(Q_{\gamma_i(2)-1}-Q_{\gamma_i(2)})}{(P_0-P_d)(Q_0-Q_d)\sum_{p=1}^{P}{1}/{E[T_p]}},\\
    &E[D(l)]\approx E[T_s^l]+E[T_s]\left(\frac{\rho}{1-\rho}\frac{c_a^2+c_s^2}{2}
    \right).\vspace{-0.1cm}
\end{split}
\end{equation}

We demonstrate in the next subsection, with our empirical results, that this is a tight lower bound.

\subsection{Simulation Results}

We describe the parameters of our distributed system with $P=5$ workers: The job arrival has a Poisson distribution with rate $0.01$. The number of tasks per matrix-vector multiplication is $1000$. We assume the time it takes for the $p$-th worker to respond to an assignment with complexity $c$ has an exponential distribution with parameter $\mu_p/c$. Here, $\mu_p$ indicates the operation rate of the $p$-th worker, and the value of $\mu_p$ for the five workers used in this section are $[350.86,591.75, 339.45,377.95,339.98]$. We consider $P_W=P_X=[8,4,0]$, and thus we have $R=d^2=4$ layers of resolution and $P_{i-1}-P_i=8$, $Q_{i-1}-Q_i=8$, for $i\in\{1,2\}$. According to Definition~\ref{def:Gamma_vec}, $\Gamma=[(1,1),(1,2),(2,1),(2,2)]$. The computational complexity of each job with no layering is set to $50$, and thus the computational complexity of each mini-job when utilizing the layering mechanism is
\begin{equation*}
    \frac{(P_{\gamma_r(1)-1}-P_{\gamma_r(1)})(Q_{\gamma_r(2)-1}-Q_{\gamma_r(2)})}{(P_0-P_d)(Q_0-Q_d)}50=\frac{16}{64}50=12.5.
\end{equation*}

\begin{figure}
    \centering
    \hspace{-0.7cm}\includegraphics[height=3.7cm,keepaspectratio]{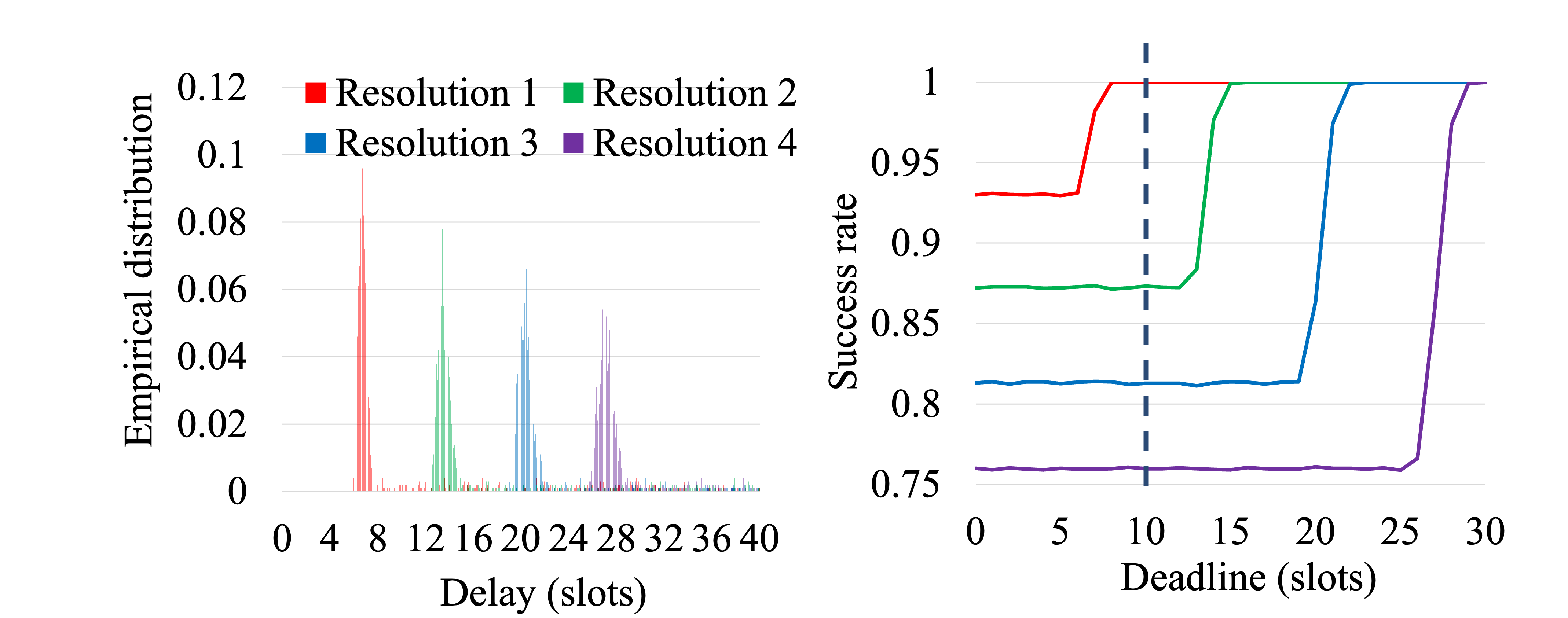}
    \caption{\small{(left) Distribution and (right) success rate, for the execution delay of $R=4$ layers of resolution, based on $1000$ jobs.}}
    \label{fig:distribution}
    \vspace{-0.3cm}
\end{figure}

Fig~\ref{fig:distribution}~(left) demonstrates the empirical delay distribution for the four layers of resolution. As seen, the higher layers have wider distributions because they accumulate the deviation from average behaviour of the earlier layers of resolution. The execution delays of different layers still stand notably far from each other across major portion of realizations. In terms of the average behaviours, the average the average delay $E[D(r)]$ and $E[D]$ for $r\in\{1,2,3,4\}$ are shown in Table~\ref{table:delay}, obtained using Monte Carlo simulation and using the theoretical lower bound derived in (\ref{eq:LB}).

\begin{table}
    \centering
    \caption{Average execution delay for various resolution layers compared to the average delay of the one-shot computation.}
    \label{table:delay}
    \begin{tabular}{|l|cccc|c|}
    \hline
        Average delay &$l=1$&$l=2$&$l=3$&$l=4$&One-shot\\
    \hline
         Empirical &$11.33$&$18.12$&$24.92$&$31.71$&$32.43$  \\
    \hline
         Theoretical LB&$10.42$&$16.67$&$22.92$&$29.17$& $29.17$ \\
    \hline
    \end{tabular}
\end{table}

We then perform an experiment where we impose a deadline to the system. The deadline specifies the maximum allowed computation time for each job when the system is busy. If the computational time of a job -- excluding the waiting time in the queue -- exceeds the deadline and there are subsequent job(s) in the queue, the job will be terminated. In the case of layering, lower resolutions of the job result might still be available although the job is terminated. The success rate is then defined as the ratio of the number of jobs, resp., certain resolution of jobs, that are finished (either because their computation time took less then the deadline or there was no other job in the queue) to the total number of jobs, for the first $1000$ jobs arrived at the queue. Fig.~\ref{fig:distribution}~(b) shows the success rate versus deadline value. As seen, for the first resolution, the success rate is $1$ for the deadline value $10$. However, the success rate for the higher resolutions or no-layering case are much lower. This experiment manifests that layering is a must-addition to distributed systems that have a deadline, to increase the effective resource utilization. Otherwise, the resources that are spent on terminated jobs will be completely wasted.

\subsection{Related Work}

In \cite{MohammadAliCodedSketch,MohammadAli2,BartanPolarSketch2023}, coding schemes are presented to recover approximated result of a distributed matrix multiplication with fewer task results than what needed for the full-precision. Although reducing the computational complexity, the presented schemes are still one-shot mechanisms.

\section{ML Classification with Adaptive Resolution}\label{sec:application_nonlinear}

In this section, through a running example, we show how the layering strategy presented in Section~\ref{sec:method}.~B can be used for evaluation of a neural network with multiple levels of resolution. In the course of this section, we will also propose an scheme with adaptive resolution for the AI-based decision-making processes.

\subsection{System Design}

\begin{figure}
    \centering
    \includegraphics[width=0.4\textwidth]{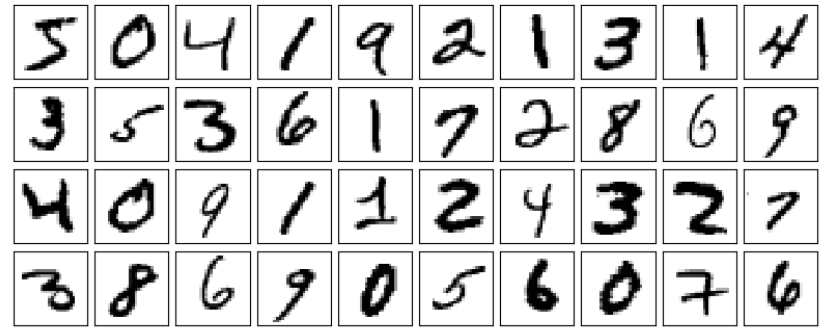}
    \caption{Several randomly-selected samples in MNIST dataset.\vspace{-0.5cm}}
    \label{fig:samples}
    \vspace{-0.4cm}
\end{figure}

We use a binary classification on MNSIT dataset as our running example to explain the solution. The MNIST dataset consists of $70{,}000$ grayscale images of handwritten digits with size $28\times 28$ \cite{deng2012mnist}. The original labels are integers in $\{0,\dots,9\}$, corresponding to the digit illustrated by each image. Fig.~\ref{fig:samples} shows several
arbitrarily-chosen samples of the MNIST dataset. The labels we use are binary, showing if the image shows an even or odd digit. The 2d samples are flattened and normalized to have scalar elements in $[0,1]$. For training, we used a neural network with $784$ input nodes, two hidden layers each having $20$ nodes, and one output. For intermediate layers and final layer, Relu and Sigmoid were used for the nonlinearity, respectively. We used SGD optimizer, batch size $100$,
learning rate $10^{-4}$, and $10$ training epochs\footnote{The focus of this paper is not training models, rather expediting their inference. Thus, both the dataset and the model architecture we use serve as a benchmark.}.\\

We present an inference technique with adaptive resolution, enabled by the layering strategy presented in this paper. Most AI-based classification solutions are associated with a threshold mechanism to assign samples to different classes. In our current classification task, a reasonable threshold is $0.5$, meaning if the output of the neural network (after the Sigmoid nonlinearity) is higher than this threshold, the sample is assigned to class even and if not, it is assigned to class odd. Most errors happen when the output of the neural network is close to the threshold. Thus, we propose upgrading the resolution only if the current resolution of the result is near the decision threshold.

\subsection{Simulation Results}

\begin{figure}
    \centering
    \includegraphics[width=0.28\textwidth]{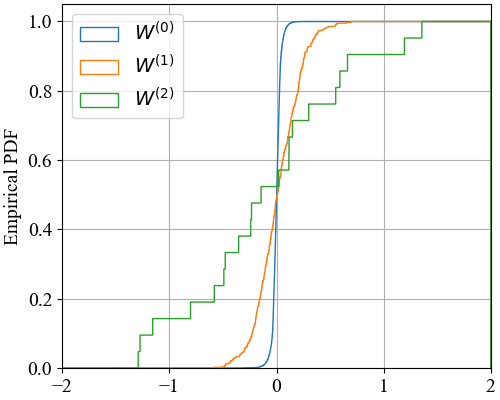}\vspace{-0.2cm}
    \caption{Model weight distribution for a deep neural network with two hidden layers that is trained to classify MNIST digits into even and odd classes.}
    \label{fig:MNIST_binary_weightdist}
    \vspace{-0.4cm}
\end{figure}

We first choose appropriate partitioning vectors for the input and the model weights. For the input samples, by design, the samples belong to the interval $[0,1]$. Thus, we choose $P_X=[0,-1,-2,-3,-4]$, and we will show later that going beyond $4$-bit budget is unnecessary for this example. As for the model weights, we benefit from the empirical distribution of the model weights in Fig.~\ref{fig:MNIST_binary_weightdist}. The cumulative distribution of the model weights indicate that $|W^{(0)}|<2^{-1}$, $
|W^{(1)}|<2^{0}$, and $|W^{(2)}|<2^{1}$. Thus, we chose the following partitioning vectors in our experiments (corresponding to $R=4$ layers of resolution):
\begin{equation*}
    \begin{split}
        P_{W,0}&=[-1,-2,-3,-4,-5],\\
        P_{W,1}&=[0,-1,-2,-3,-4],\\
        P_{W,2}&=[1,0,-1,-2,-3],
    \end{split}
\end{equation*}
which correspond to one additional bit per resolution upgrade.

\begin{figure}
    \centering
    \begin{tabular}{cc}
      \includegraphics[height=4.8cm]{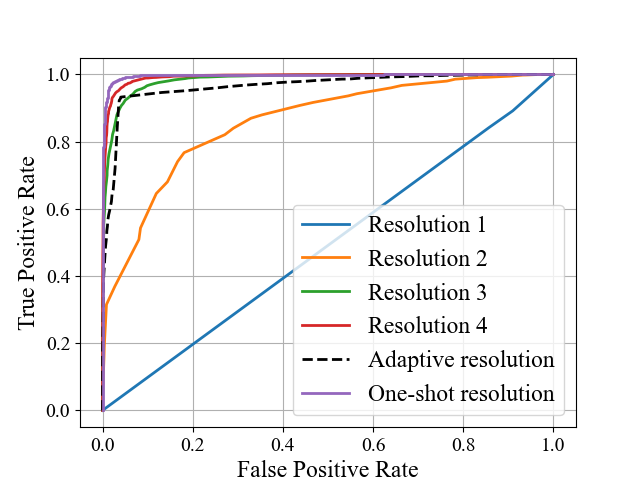}\hspace{-01.05cm}   &
      \includegraphics[height=5cm]{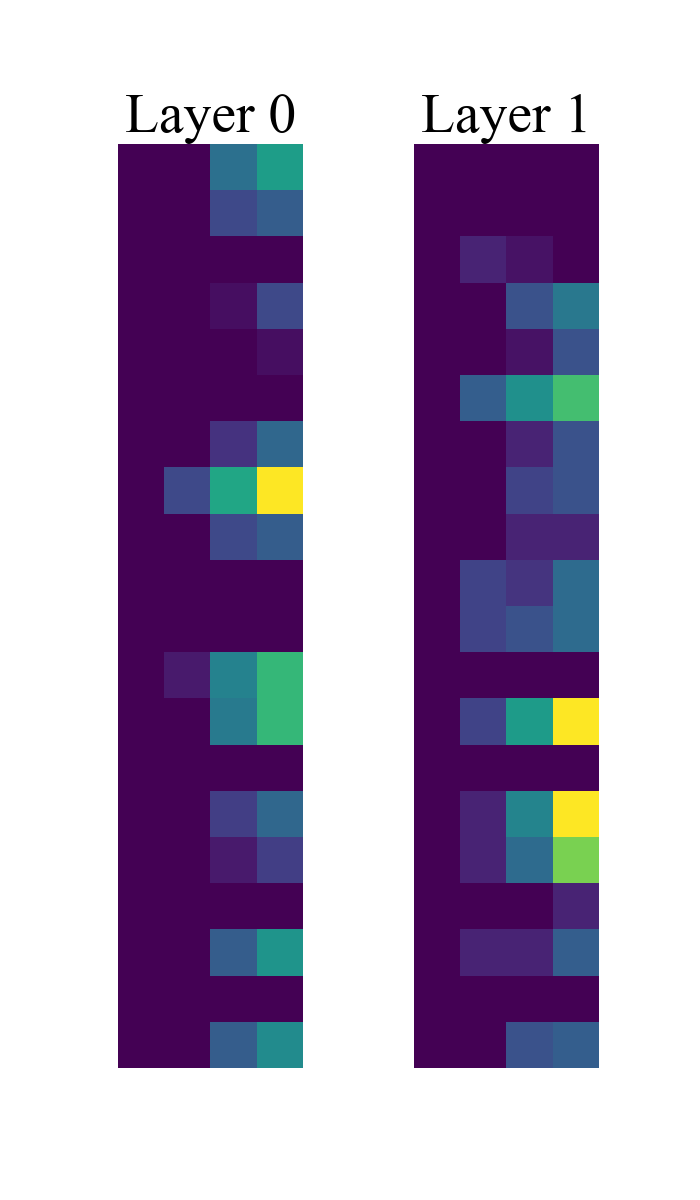}\\
      (a)   & (b)
    \end{tabular}
    \caption{(a) ROC curves for one-shot computation versus different resolution upgrades. (b) Values of the hidden nodes after applying the nonlinearity at different resolution upgrades of a sample.}
    \label{fig:MNIST_binary_ROC_Activations}
    \vspace{-0.4cm}
\end{figure}

We then use Algorithm~\ref{alg:main_algo} to obtain the inference result with $R=4$ resolution upgrades per inference. We use $h_\text{min}=-4$ for this example, as we empirically found it to be sufficient for representing $H_r^{(l)}$.
Fig.~\ref{fig:MNIST_binary_ROC_Activations}~(a) shows the ROC curves for the one-shot inference and for the four levels of the layering strategy. Table~\ref{table:mnist_binary_result} shows the ratio of correct predictions and the area under the ROC curve (ROC-AUC). As we see the prediction accuracy starts to go up after the second level of resolution, and after the forth resolution upgrade, the final inference performance is very close to the one-shot computation.

\begin{table}
    \centering
    \caption{Correct prediction ratio and ROC AUC for one-shot computation versus different resolutions of the presented layering strategy for DNN evaluations.}
    \begin{tabular}{|c|c|c|}
    \hline
         Inference strategy & Correct prediction ratio & ROC AUC\\
         \hline
         Resolution 1& $0.5074$ & $0.4918$\\
         Resolution 2& $0.7694$ & $0.8587$\\
         Resolution 3& $0.9188$ & $0.9839$\\
         Resolution 4& $0.9366$ & $0.9922$\\
         One-shot computation& $0.9762$ & $0.9968$\\
        Adaptive resolution& $0.9347$ & $0.9654$\\
        \hline
    \end{tabular}
    \label{table:mnist_binary_result}
    \vspace{-0.2cm}
\end{table}

The reason behind why the prediction performance is low for the first resolution can be explained through Fig.~\ref{fig:MNIST_binary_ROC_Activations}~(b). This figure shows the color-coded value of each hidden node (after applying the nonlinearity), at various inference resolution of one sample image. Overall, we have 20 nodes at the first hidden layer and 20 nodes at the second hidden layer. The color of each node indicates whether the node is active (colored) or not (black). We observe that, at the first resolution upgrade, all nodes at intermediate layer 1 have value zero (are inactive), leading to inactive nodes at intermediate layer 2 as well. This is because the only bit of elements in $W^{(1)}$ that are incorporated in the first resolution upgrade has positional value $2^{-1}$. There are only a small percentage of the weights that have non-zero value at this position, and thus $W^{(1)}$ is almost zero in the first resolution upgrade. This observation suggests to incorporate more number of bits at the first resolution upgrade.

\begin{figure}
    \centering
    \begin{tabular}{cc}
         \includegraphics[height=3.2cm]{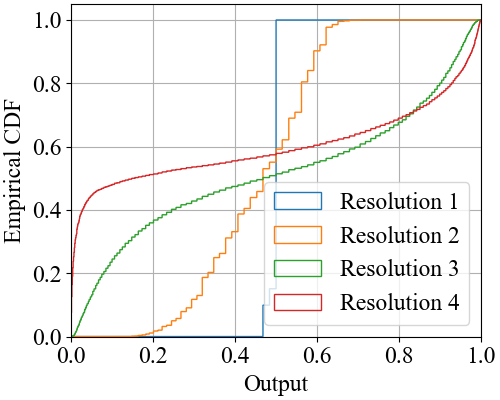}\hspace{-3cm}&
         \includegraphics[height=3.5cm]{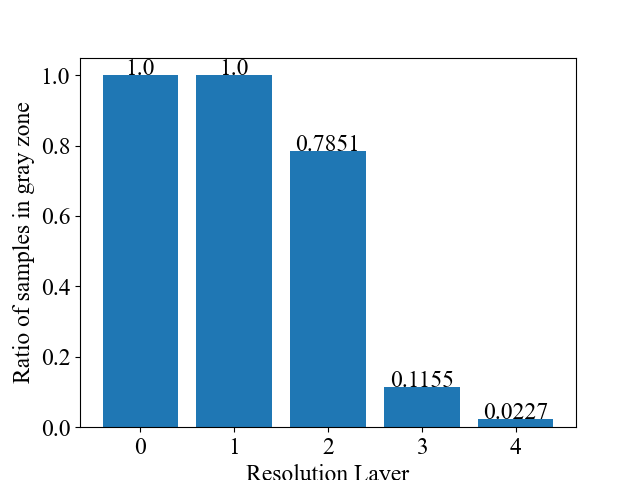}\\
         (a)&(b)
    \end{tabular}

    \caption{(a) CDF of the output at different layers of resolution. (b) The ratio of samples that remain in the gray zone and demand a resolution upgrade at the end of each computational resolution.\vspace{-0.2cm}}
    \label{fig:MNIST_adaptive}
    \vspace{-0.2cm}
\end{figure}

Fig.~\ref{fig:MNIST_adaptive}~(a) shows the empirical distribution of the output of the neural network for $10{,}000$ test samples obtained at different resolution upgrades. We observe that for the first resolution upgrade, almost the entire test set have outputs close to the threshold. For the second resolution upgrade, the outputs are almost uniformly spread across the interval $[0.2,0.7]$, and for higher resolution upgrades, most samples have output values that stand far from the threshold. This observation and the possibility of having layered resolution we introduced in this paper motivated us to implement an adaptive mechanism to just perform the resolution upgrade for a sample if the current output is close to the threshold, i.e., the output is in a so-called \textit{gray zone} set to $[0.3,0.6]$ here.

The amount of computational resources that are saved thanks to the presented adaptive strategy is manifested in Fig.~\ref{fig:MNIST_adaptive}~(b). At first, there are no-prior information about output of samples and they are all in gray zone. After the first resolution upgrade, still all samples need resolution upgrade. However, after the second resolution upgrade (resolution $2$), around $79\%$ of samples need further resolution upgrade. This ratio goes down after the third resolution upgrade, and only $12\%$ percent of samples require the forth resolution upgrade. The performance of this adaptive strategy is shown in the ROC curve in Fig.~\ref{fig:MNIST_binary_ROC_Activations}~(a) and in Table~\ref{table:mnist_binary_result}. In terms of the predictive utility, the performance of adaptive strategy is very close to the case where all samples go through all resolution upgrades.

\subsection{Related Work}

The computational cost of the ML inference task can be significantly large and reducing this cost has recently attracted notable attention. The previous studies in this area can be categorized into two groups: Static methods (also known as model compression) and dynamic methods (also called dynamic inference). In model compression, the model architecture is modified to attain lightweight networks and thus have less inference cost. Methods such as network pruning and low-bit quantization are well-known techniques under this umbrella. Network pruning aims to prune the less important parts of the model, e.g,  \cite{li2017pruning,10.5555/3157096.3157329,9180094,pmlr-v139-wang21e,Li_2021_CVPR}. In low-bit quantization, the learned model weights are represented using few bits instead of high precision floating points, e.g., \cite{10.5555/3122009.3242044,10.1007/978-3-319-46493-0_32, 10.5555/3157382.3157557,Jing_2021_ICCV,10.5555/3045118.3045303,Jacob_2018_CVPR}.

The dynamic inference includes strategies that adjust the complexity of the model depending on the complexity of each sample. The main idea is that not all inputs require the same amount of computation to yield a confident prediction. By dynamic inference, the model architecture and the sample representation can be modified depending on each sample complexity \cite{AdaptiveInference1}. For instance, \cite{NEURIPS2021_2d969e2c,NEURIPS2019_bd853b47,NEURIPS2021_64517d84} propose exploiting the spatial redundancy of input samples to reduce the computational costs, and in \cite{Sun_2021_ICCV,Ghodrati_2021_CVPR,DBLP:conf/nips/ZhuHWZNLW21,9156745}  lightweight mechanisms are proposed to identify which numerical precision to be used for each sample. In \cite{9157556,ZhangASPLOS2021,9157556,Yang_2020_CVPR}, several sub-models are jointly trained, each for a specific resolution of the model weights and data samples. Then, this sub-models can be used, either from simplest to the highest needed precision or be selected depending on the resource budgets at the time of inference.

Early exit is another dynamic inference technique where a series of models are trained, and then a sample can exit the trained model earlier depending on the  confidence of the current result  \cite{10.1145/3394171.3413701,9156492,DistillationPhuong2019,Teerapittayanon2016BranchyNetFI, 10.1145/3469116.3470012,PASSALIS2020107346,10.1145/3400302.3415698}. The subsequent model for early exit can also be replaced with hierarchical tree structure \cite{Ruiz_Verbeek_2021}, or by multiple classifiers \cite{huang2018multiscale} to allocate uneven computation across different samples. A similar technique is to selectively include or remove certain parts of the model on a per-input basis, aiming to improve inference efficiency \cite{Wang_2018_ECCV_SkipNet,9009445,Meng_2022_CVPR,Xue_Li_Zhang_2020,blockdrop,9345986,Cai_2021_WACV,9157150}.

The adaptive-resolution technique presented in this section can be considered as a dynamic inference technique since it offers different complexity per sample depending on the intermediate results. However, it is fundamentally different from the state-of-the-art methods in this group, as our layering strategy does not require changing the model architecture or training. Moreover, the same model can be used for each resolution upgrade, as the result of each resolution upgrade is an indicator of the current prediction confidence. This feature mitigates the accuracy degradation and the restrictions imposed on the model architecture, presented in the previous dynamic inference methods.

\section{Discussion and Future Work}\label{sec:discussion}

In this paper, we proposed strategies to perform massive computational algorithms in successive layers of resolution. Each resolution layer has a computational complexity smaller than the one for the one-shot computation, and also the result of the final layer is close to the one-shot result. Our strategies are for the linear transformations and also for the evaluation of a deep neural network. We believe the result of this work can bring more adaptability to the massive computational systems, as the user can have intermediate results, observe how the computations are going, and decide if higher resolutions are needed. Besides, the presented solutions can be beneficial for deadline-based systems, as early results for terminated jobs are preferred over having no result. The presented mechanisms as discussed can be incorporated into the cloud solutions and also into the schedulers designed for the ML algorithms. As for future work, extending the layering mechanisms for other ML architectures, such as transformers and convolutional networks, are promising research directions. Moreover, incorporating the adaptive mechanism for ML training to reduce the training cost is an impactful direction to pursue. In the layering techniques we discussed in this paper, resolution upgrades are successively added together to reflect the refined result. Having successive refinement techniques that are not necessarily additive is another future research direction.


\appendix\label{appendix:proofs}

\noindent\textbf{Proof of Theorem~\ref{theorem:valid_linear1}:} Given the partitioning vectors $P_W=[P_0,\dots,P_d]$ and $P_X=[Q_0,\dots,Q_d]$, we have $W=\sum_{i=1}^{d} W_i 2^{P_i}+ R_W$ and $X=\sum_{j=1}^{d} X_j 2^{Q_j} + R_X$, where $|R_W|<2^{P_d}$ and $|R_X|<2^{Q_d}$. Thus,
\begin{equation*}
\begin{split}
f(X)&=WX=\sum_{k=1}^{R}W_{\gamma_k(1)}X_{\gamma_k(2)} 2^{P_{\gamma_k(1)}+Q_{\gamma_k(2)}}+R_f,\\
    R_f&=R_X \sum_{i=1}^{d} W_i 2^{P_i} + R_W\sum_{j=1}^{d} X_j 2^{Q_j} + R_WR_X.
\end{split}
\end{equation*}
Since $|\sum_{i=1}^{d} W_i 2^{P_i}|< 2^{P_0}$ and $|\sum_{j=1}^{d} X_j 2^{Q_j}|< 2^{Q_0}$, using triangle inequality, we have
\begin{equation*}
    |R_f| < 2^{P_0+Q_d} + 2^{P_d+Q_0} + 2^{P_d+Q_d}.
\end{equation*}
On the other hand, by definition, for $r\in\{1,\dots,R\}$,
\begin{equation*}
\Omega_r=\sum_{i=1}^{r}W_{\gamma_i(1)}X_{\gamma_i(2)} 2^{P_{\gamma_i(1)}+Q_{\gamma_i(2)}}.
\end{equation*}
Thus,
\begin{equation*}
    \begin{split}
        |f(X)-\Omega_r|&=|R_f+\sum_{i=r+1}^RW_{\gamma_i(1)}X_{\gamma_i(2)} 2^{P_{\gamma_i(1)}+Q_{\gamma_i(2)}}|\\
        &<|R_f|+|\sum_{i=r+1}^RW_{\gamma_i(1)}X_{\gamma_i(2)} 2^{P_{\gamma_i(1)}+Q_{\gamma_i(2)}}|\\
        &<|R_f| +(R-r)2^{P_{\gamma_r(1)}+Q_{\gamma_r(2)}}.
    \end{split}
\end{equation*}
By Definition~\ref{def:Gamma_vec}, $P_{\gamma_r(1)}+Q_{\gamma_r(2)}$ is non-increasing in $r$, and $(R-r)$ is decreasing in $r$. Thus, $\delta(r)=\delta +(R-r)2^{P_{\gamma_r(1)}+Q_{\gamma_r(2)}}$ is non-increasing in $r\in\{1,\dots,R\}$, and
\begin{equation*}
    \delta=2^{P_0+Q_d} + 2^{P_d+Q_0} + 2^{P_d+Q_d}.
\end{equation*}
\qed

\noindent\textbf{Proof of Theorem~\ref{theorem:valid_linear2}:} First, we study the computational complexity of $f(X)=WX$, where $W$ is a $u\times v$ matrix and $X$ is a $v\times 1$ vector. The partitioning vectors for $W$ and $X$ are defined as $P_W=[P_0,\dots,P_d]$ and $P_X=[Q_0,\dots,Q_d]$, respectively. For simplicity, we assume the bits that have positional values lass than $P_d$ (resp., $Q_d$) can be neglected for $W$ (resp., $X$). Thus, each element of $W$ has $P_0-P_d$ important bits and each element of $X$ has $Q_0-Q_d$ important bits. As a result,
\begin{equation*}
    \Theta(f)\approx uv (P_0-P_d)(Q_0-Q_d).
\end{equation*}
Next, we remind that for $r\in\{1,\dots,R\}$,
\begin{equation*}
    f_r(X,\Omega_{r-1})=\Omega_{r-1}+W_{\gamma_r(1)}X_{\gamma_r(2)} 2^{P_{\gamma_r(1)}+Q_{\gamma_r(2)}},
\end{equation*}
The tuples $(\gamma_r(1),\gamma_r(2))$ are defined in Definition~\ref{def:Gamma_vec}, and $\cup_{r=1}^R (\gamma_r(1),\gamma_r(2))=\{1,\dots,d\}^2$. Computing $f_r(X,\Omega_{r-1})$ requires multiplying a $u\times v$ matrix with $(P_{\gamma_r(1)-1}-P_{\gamma_r(1)})$-bit  elements with a $v\times 1$ vector with $(Q_{\gamma_r(2)-1}-Q_{\gamma_r(2)})$-bit  elements, and its complexity is
\begin{equation*}
    \Theta(f_r)= uv (P_{\gamma_r(1)-1}-P_{\gamma_r(1)})(Q_{\gamma_r(2)-1}-Q_{\gamma_r(2)}).
\end{equation*}
Therefore,
\begin{equation*}
\begin{split}
    \sum_{r=1}^R\Theta(f_r) &= uv \sum_{r=1}^R (P_{\gamma_r(1)-1}-P_{\gamma_r(1)})(Q_{\gamma_r(2)-1}-Q_{\gamma_r(2)})\\
    &=uv\sum_{i=1}^{d}\sum_{j=1}^d (P_{i-1}-P_{i})(Q_{j-1}-Q_{j})\\
    &=uv\sum_{i=1}^{d}(P_{i-1}-P_{i})\sum_{j=1}^d(Q_{j-1}-Q_{j})\\
    &=uv\left(P_0-P_d\right)\left(Q_0-Q_d\right).
\end{split}
\end{equation*}
\qed

\noindent\textbf{Proof of Theorem~\ref{theorem:valid_nonlinear1}:} Let input and weights associated the $r$-th resolution upgrade be $X_r$ and $\{W_r^{(0)},\dots,W_r^{(L)}\}$. The input and weights corresponding to the one-shot result are $X_r+\Delta X_r$, and $\{W_r^{(0)}+\Delta W_r^{(0)},\dots,W_r^{(L)}+\Delta W_r^{(L)}\}$. We denote with $H_r^{(l)}$ and $H^{(l)}$ the values related to the $l$-th hidden layer of the model, at the $r$-th resolution and at the full-resolution (one-shot result), respectively, where $l\in\{0,\dots,L+1\}$. Here, we assume $|H_r^{(l)}|,|H^{(l)}|<2^{h_\text{max}}$.\footnote{The combination of finite initializations, activation functions, and regularization techniques helps ensure that the hidden nodes of a neural network take finite values during both the forward pass and the training process.} We define $\Delta H_r^{(l)}=H^{(l)}-H_r^{(l)}$, $r\in\{1,\dots,R\}$. According to Lemma~\ref{lemma:distortion}, $\Delta H_r^{(0)}=\Delta X_r$,
\begin{equation}\label{eq:expanding_inference_error}
\begin{split}
    &\Delta H_r^{(l)}=W_r^{(l-1)}\sigma_{l-1}^{H_r^{(l-1)}}(\Delta H_r^{(l-1)})\\
    &{+}\Delta W_r^{(l-1)}\sigma_{l-1}(H_r^{(l-1)}){+}\Delta W_r^{(l-1)}\sigma_{l-1}^{H_r^{(l-1)}}(\Delta H_r^{(l-1)}),
\end{split}
\end{equation}
and $f(X)= \sigma'(H_r^{(L+1)}+\Delta H_r^{(L+1)})$. The last term in (\ref{eq:expanding_inference_error}) is much smaller compared to the other two terms and can be neglected. We then apply the triangle inequality,
\begin{equation}\label{eq:expanding_inference_error2}
\begin{split}
    |\Delta H_r^{(l)}|&<|W_r^{(l-1)}\sigma_{l-1}^{H_r^{(l-1)}}(\Delta H_r^{(l-1)})|\\
    &+ |\Delta W_r^{(l-1)}\sigma_{l-1}(H_r^{(l-1)})|.
\end{split}
\end{equation}

We first characterize the term $\sigma_{l-1}^{H_r^{(l-1)}}(\Delta H_r^{(l-1)})$ in (\ref{eq:expanding_inference_error2}). According to Lemma~\ref{lemma_nonlinearity},
\begin{equation*}
    \sigma_{l-1}^{H_r^{(l-1)}}(\Delta H_r^{(l-1)})=\Gamma \Delta H_r^{(l-1)}+\Lambda H_r^{(l-1)}+V\approx \Gamma \Delta H_r^{(l-1)}.
\end{equation*}
The variable $\Gamma$ is a diagonal matrix, whose elements indicate the slop of the piece-wise linear function at various regions depending on  the elements of  $H_r^{(l-1)}$. The variables $\Lambda$ and $V$ represent the effect of discrepancy between the region of the function $\sigma_l$ when applied to $H_r^{(l-1)}$ versus when applied to $H^{(l-1)}$. After a few resolution upgrades, this discrepancy vanishes and we assume $\Lambda=[0]$ and $V=[0]$ when $r>1$ for well-behaved partitioning vectors. Therefore,
\begin{equation*}
    |\Delta H_r^{(l)}|{<}|W_r^{(l-1)}\Gamma^{(l-1)} \Delta H_r^{(l-1)}|+ |\Delta W_r^{(l-1)}\Gamma^{(l-1)} H_r^{(l-1)}|.
\end{equation*}

The partitioning vector $P_{W,l-1}=[P_0^{(l-1)},\dots,P_R^{(l-1)}]$ requires $|W_r^{(l-1)}|<2^{P_0^{(l-1)}}$. We also remind that elements of the diagonal matrix $\Gamma^{(l-1)}$ are non-negative and less then one.\footnote{The derivative of the practical nonlinear activation functions such as leaky Relu, Sigmoid, and Tanh is less than one. Thus, their piece-wise linear approximations have coefficients less than one.} Besides, $|\Delta W_r^{(l-1)}|<2^{P_{r}^{(l-1)}}$, leading to
\begin{equation*}
    |\Delta H_r^{(l)}|{<}2^{P_0^{(l-1)}}n_{l-1}|\Delta H_r^{(l-1)}|+2^{h_\text{max}}n_{l-1}2^{P_r^{(l-1)}}[1]_{n_l}
\end{equation*}
By denoting an element of $\Delta H_r^{(l)}$ as $\Delta h_r^{(l)}$, we obtain
\begin{equation}\label{eq:summarybeforelemma}
    |\Delta h_r^{(l)}|{<}2^{P_0^{(l-1)}}n_{l-1}|\Delta h_r^{(l-1)}|+2^{h_\text{max}}n_{l-1}2^{P_r^{(l-1)}}.
\end{equation}

We next bring a lemma that will be used for the rest the proof.
\begin{lemma}\label{lemma:helper_precision_inference}
    Consider non-negative scalars $a_l$ and $b_l$, for $l\in\{0,1,2,\dots\}$, and $\theta_0\geq 0$. Given $\theta_l<a_{l-1}+b_{l-1}\theta_{l-1}$, we have
    \begin{equation*}
        \theta_l<\theta_0\prod_{i=0}^{l-1} b_i + \sum_{i=0}^{l-1}\left(a_i\prod_{j=i+1}^{l-1}b_j\right),
    \end{equation*}
\end{lemma}
for $l\in\{1,2,\dots\}$.
\begin{proof}
    We prove the lemma by induction. For $l=1$, $\theta_1<a_{0}+b_{0}\theta_{0}$ checks out. We assume the following is correct
    \begin{equation*}
        \theta_{l-1}<\theta_0\prod_{i=0}^{l-2} b_i + \sum_{i=0}^{l-2}\left(a_i\prod_{j=i+1}^{l-2}b_j\right).
    \end{equation*}
    Thus,
    \begin{equation*}
        \begin{split}
            \theta_l&<a_{l-1}+b_{l-1}\theta_{l-1}\\
            &<
            a_{l-1}+b_{l-1}\left(\theta_0\prod_{i=0}^{l-2} b_i + \sum_{i=0}^{l-2}\left(a_i\prod_{j=i+1}^{l-2}b_j\right)\right)\\
            &=\theta_0\prod_{i=0}^{l-1} b_i+a_{l-1}+\sum_{i=0}^{l-2}\left(a_i\prod_{j=i+1}^{l-1}b_j\right)\\
            &=\theta_0\prod_{i=0}^{l-1} b_i + \sum_{i=0}^{l-1}\left(a_i\prod_{j=i+1}^{l-1}b_j\right),
        \end{split}
    \end{equation*}
    and the lemma proof is complete.
\end{proof}
Using Lemma~\ref{lemma:helper_precision_inference} to simplify (\ref{eq:summarybeforelemma}), we obtain
\begin{equation}\label{eq:delta_h}
\begin{split}
    |\Delta& h_r^{(l)}|<|\Delta h_r^{(0)}|\prod_{i=0}^{l-1} 2^{P_0^{(i)}}n_{i} \\
    &+ \sum_{i=0}^{l-1}\left(
    2^{h_\text{max}}n_{i}2^{P_r^{(i)}}
    \prod_{j=i+1}^{l-1}2^{P_0^{(j)}}n_{j}\right).
\end{split}
\end{equation}
Considering $\Delta H_r^{(0)}=\Delta X_r$ and using the partitioning matrix $P_X=[Q_0,\dots,Q_R]$, we have $|\Delta h_r^{(0)}|<2^{Q_{r}}$. Finally,
\begin{equation*}
\begin{split}
    |f&(X)-\Omega_r|=|\sigma'(H_r^{(L+1)}+\Delta H_r^{(L+1)})-\sigma'(H_r^{(L+1)})|\\
    &\approx |J(H^{(L+1)})\Delta H_r^{(L+1)})|<|J(H^{(L+1)})||\Delta H_r^{(L+1)})|.
\end{split}
\end{equation*}
Here, $J(H^{(L+1)})$ is the Jacobian matrix of the final activation function $\sigma'(.)$ at $H^{(L+1)}$, and it is bounded in practice.
For example, for the Sigmoid function, $|J(H^{(L+1)})|<0.25$. Thus,
\begin{equation}\label{eq:Jacobian}
    |f(X)-\Omega_r|<J_\text{max}m |\Delta h^{(L+1)}|[1]_m.
\end{equation}
By combining (\ref{eq:delta_h}) and (\ref{eq:Jacobian}), and separating the terms that have dependencies on $r$, we obtain
\begin{equation*}
\begin{split}
    |f(X)-\Omega_{r}|&<2^{Q_{r}}\left(J_\text{max}m\prod_{i=0}^{l-1} 2^{P_0^{(i)}}n_{i}\right) \\
    &+ \sum_{i=0}^{l-1}2^{P_r^{(i)}}\left(
    J_\text{max}m2^{h_\text{max}}n_{i}
    \prod_{j=i+1}^{l-1}2^{P_0^{(j)}}n_{j}\right).
\end{split}
\end{equation*}
The terms in parenthesis are non-negative and independent of $r$ and the terms prior to the parentheses are non-increasing with respect to $r$. Thus, the overall upper bound that represents $\delta(r)$ is non-increasing in $r$.
\qed

\noindent\textbf{Proof of Theorem~\ref{theorem:valid_nonlinear2}:}
We identify the computational complexity of Algorithm~\ref{alg:main_algo} for each resolution upgrade and compare it with the one-shot inference complexity. The partitioning vectors for $W^{(l)}$ and $X$ are given as $P_{W,l}=[P_0^{(l)},\dots,P_R^{(l)}]$ and $P_X=[Q_0,\dots,Q_R]$, respectively. For simplicity, we assume the bits that have positional values lower than $P_R^{(l)}$ (resp., $Q_R$) can be neglected for $W^{(l)}$ (resp., $X$). The complexity of bit shifting operation is negligible, and we also assume the computational complexity of addition is negligible compared to the multiplication. The computational complexity of piece-wise linear functions are considered as $\Theta(\sigma)=1$ and $\Theta(\sigma')=m$ for each of their outputs. Thus, the computational complexity are mainly due to the lines \ref{algo2:17}, \ref{algo2:14}, and \ref{algo2:21}.

Since matrix $H$ has $(h_\text{max}-h_\text{min})$-bit elements, it is fair to assume $\Delta H$ has elements with at most $0.5(h_\text{max}-h_\text{min})$ effective bits. At line~\ref{algo2:14} for $l\in\{0,\dots,L\}$, we need to perform three matrix multiplications: First one is $A_l \Delta H_\sigma$, which requires multiplying an $n_{l+1}\times n_{l}$ matrix with $(P_0^{(l)}-P_r^{(l)})$-bit elements and a vector of size $n_l$ with $0.5(h_\text{max}-h_\text{min})$-bit elements. Second one is $\Delta A_l \sigma_l(H)$, which requires multiplying an $n_{l+1}\times n_{l}$ matrix with $(P_{r-1}^{(l)}-P_r^{(l)})$-bit elements and a vector of size  $n_l$ with $(h_\text{max}-h_\text{min})$-bit elements, plus an additional $n_{l}$ times overhead due to applying the point-wise function $\sigma_l$. Third one is $\Delta A_l \Delta H_\sigma $, which requires multiplying an $n_{l+1}\times n_{l}$ matrix with $(P_{r-1}^{(l)}-P_r^{(l)})$-bit elements and a vector of size  $n_l$ with $0.5(h_\text{max}-h_\text{min})$-bit elements. At line \ref{algo2:17} and \ref{algo2:21}, we have additional $n_{l+1}$ and $m^2$ units of computational overhead, respectively. Thus, the complexity of the $r$-th resolution upgrade is bounded as follows:
\begin{equation*}
    \begin{split}
        \Theta(f_r)&<\sum_{l=0}^{L} n_{l}n_{l-1}(h_\text{max}{-}h_\text{min})(0.5P_0^{(l)}{+}1.5P_{r-1}^{(l)}{-}2P_r^{(l)})
        \\&+2\sum_{l=0}^{L} n_{l}+m^2
    \end{split}
\end{equation*}

On the other hand, the computational complexity of the one-shot result is,
\begin{equation*}
    \begin{split}
    \Theta(f)&=\sum_{l=0}^{L} n_{l}n_{l-1}(h_\text{max}{-}h_\text{min})(P_0^{(l)}-P_R^{(l)})+\sum_{l=0}^{L} n_{l}+m^2
    \end{split}
\end{equation*}
Therefore, $\Theta(f){-}\Theta(f_r)>\epsilon(r)$, and
\begin{equation}\label{eq:comp_th4}
\begin{split}
    \epsilon(r)&{=}\hspace{-0.1cm}\sum_{l=0}^{L}\hspace{-0.1cm} n_{l}n_{l-1}(h_\text{max}{-}h_\text{min})(0.5P_0^{(l)}{-}1.5P_{r-1}^{(l)}{+}2P_r^{(l)}{-}P_R^{(l)})
\end{split}
\end{equation}
Here, a term $\sum_{l=0}^{L} n_{l}$ is ignore because it is negligible compared to the other terms. We can rewrite (\ref{eq:comp_th4}) as follows,
\begin{equation}\label{eq:comp_inf}
    \begin{split}
        \epsilon(r)&=0.5\sum_{l=0}^{L} n_{l}n_{l-1}(h_\text{max}{-}h_\text{min})(P_0^{(l)}{-}P_R^{(l)})\\
        &+0.5\sum_{l=0}^{L} n_{l}n_{l-1}(h_\text{max}{-}h_\text{min})(P_r^{(l)}{-}P_R^{(l)})\\
        &-1.5\sum_{l=0}^{L} n_{l}n_{l-1}(h_\text{max}{-}h_\text{min})(P_{r-1}^{(l)}{-}P_{r}^{(l)})
    \end{split}
\end{equation}
In (\ref{eq:comp_inf}), all terms in the summations are non-negative, because a partitioning vector has decreasing elements. Moreover, the last term is much smaller then the first two terms since the difference between adjacent elements of a partitioning vector is smaller than the one for elements that are farther away. Thus, $\epsilon(r)>0$.
\qed

\off{\section*{Acknowledgment}
This work has been supported by the MIT Portugal Program, the National Science Foundation under grant no. CNS-2008624, and the European Union's Horizon 2020 Research And Innovation Programme under grant no. 694630.}

\bibliographystyle{IEEEtran}
\bibliography{IEEEabrv,references_short}

\end{document}